\documentclass[
aip, % aps [default], aip, aapm, sor
jap, 
reprint, % onecolumn, preprint, reprint,
10pt, % [12pt for preprint, 10pt for twocolumn]
superscriptaddress, % groupedaddress [default], superscriptaddress, unsortedaddress, runinaddress,
amssymb, % noamssymb,
amsmath, % noamsfonts,
showkeys, % noshowkeys
floats, % endfloats, endfloats*
final, 
letterpaper, % letterpaper [default], a4paper, a5paper, oneside, twoside
balancelastpage, % balancelastpage [default for twocolumn], nobalancelastpage
flushbottom, % flushbottom [default for twocolumn], raggedbottom,
citeautoscript,
]{revtex4-2}
\usepackage{hyperref}
\usepackage{url}
\usepackage{bm}
\usepackage{overpic}
\usepackage{pict2e} % for drawing polygons
\graphicspath{{./}}

\usepackage{longtable}
\usepackage{multirow}
\usepackage{helvet} % for sans serif

\usepackage{xcolor}

\newcommand{\rv}[1]{{\color{black}{#1}}}
\makeatletter
\def\@email#1#2{%
 \endgroup
 \patchcmd{\titleblock@produce}
  {\frontmatter@RRAPformat}
  {\frontmatter@RRAPformat{\produce@RRAP{*#1\href{mailto:#2}{#2}}}\frontmatter@RRAPformat}
  {}{}
}%
\makeatother

\begin{document}
\title{Control of ferroelectric domain wall dynamics by point defects: Insights from ab initio based simulations}
\author{Sheng-Han Teng}
\email{sheng-han.teng@rub.de}
\author{Aris Dimou}
\author{Benjamin Udofia}
\author{Majid Ghasemi}
\author{Markus Stricker}
\author{Anna Grünebohm}
\email{anna.gruenebohm@rub.de}
\affiliation{Interdisciplinary Centre for Advanced Materials Simulation (ICAMS) and \\ Center for Interface-Dominated High Performance Materials (ZGH), Ruhr-University Bochum, Germany}
\date{\today) (Accepted by Journal of Applied Physics. DOI: \texttt{10.1063/5.0259824}}

\begin{abstract}
The control of ferroelectric domain walls and their dynamics on the nanoscale becomes increasingly important for advanced nanoelectronics and novel computing schemes. 
One common approach to tackle this challenge is the pinning of walls by point defects. 
The fundamental understanding on how different defects influence the wall dynamics is, however, incomplete. 
In particular, the important class of defect dipoles in acceptor-doped ferroelectrics is currently underrepresented in theoretical work. 
In this study, we combine molecular dynamics simulations based on an \textit{ab\ initio} derived effective Hamiltonian and methods from materials informatics, and analyze the impact of these defects on the motion of 180$^{\circ}$ domain walls in tetragonal BaTiO$_3$. 
We show how these defects can act as local pinning centers and restoring forces on the domain structure. 
Furthermore, we reveal how walls can flow around sparse defects by nucleation and growth of dipole clusters, and how pinning, roughening and bending of walls depend on the defect distribution. 
Surprisingly, the interaction between acceptor dopants and walls is short-ranged. 
We show that the limiting factor for the nucleation processes underlying wall motion is the defect-free area in front of the wall.
\end{abstract}
\maketitle

%%%%%%%%%%%%%%%%%%%%%%%%%%%%%%

\section{Introduction}
The success of ferroelectric materials in a plethora of applications depends on reversible polarization ($\bm{P}$) switching by electrical fields ($\bm{E}$), either at the macroscopic level or at the nanoscale. \cite{whatmore100YearsFerroelectricity2021, grunebohmInterplayDomainStructure2021}
Particularly the latter is related to domain walls (DWs), i.e.\ the interfaces that separate domains with different polarization directions.
These nanoscale objects move in applied fields and offer tremendous potential for miniaturization and realization of advanced nanoelectronics and novel computing schemes, such as neuromorphic computing. \cite{meierFerroelectricDomainWalls2022, sunNonvolatileFerroelectricDomain2022, sharmaNeuromorphicFunctionalityFerroelectric2023} 
Moreover, the motion of these walls or wall segments contributes significantly to the macroscopic dielectric \cite{jesseDirectImagingSpatial2008,mokryIdentificationMicroscopicDomain2017} and piezoelectric \cite{damjanovicLogarithmicFrequencyDependence1997, jesseDirectImagingSpatial2008, fancherContribution180degDomain2017} properties of ferroelectric materials. 
The full potential of DWs can however only be tapped by their macroscopic and nanoscale control.

Due to their importance, ferroelectric DWs and their dynamics were already studied by Merz in the early 1950s.
It was shown that the DW velocity increases with temperature ($T$) and field strength ($E$) proportional to $\exp(-1/ET)$. \cite{merzDomainFormationDomain1954}
Furthermore, it was reported early that ferroelectric DWs are Ising-type and atomically sharp.\cite{grunebohmInterplayDomainStructure2021}
In recent years, these walls have been revisited and it is now consensus that wall motion is indeed a thermally activated process even in defect-free materials,
as the walls are pinned by the underlying lattice.\cite{chongThermalActivationFerroelectric2008, klompThermalStabilityNanoscale2022, durdievDeterminingThermalActivation2024,paruchNanoscaleStudiesDomain2006, grunebohmDomainStructureTetragonal2012, padillaFirstprinciplesInvestigation180deg1996, bauerStructuresVelocitiesNoisy2022}
Microscopic simulations revealed that wall motion is initiated by the nucleation and growth of two-dimensional (2D) clusters in front of the wall \cite{shinNucleationGrowthMechanism2007, bodduMolecularDynamicsStudy2017, khachaturyanDomainWallAcceleration2022, klompThermalStabilityNanoscale2022} and is thus governed by local switching events rather than the macroscopic momentum of the walls.
The classical interpretation of walls and their motion has been challenged by reports on Bloch and N\'eel-like polarization rotation \cite{leeMixedBlochNeelIsingCharacter2009, weiNeellikeDomainWalls2016, dhakaneGraphDynamicalNeural2023, zatterinAssessingUbiquityBloch2024} and complex multi-step switching on DWs.\cite{genenkoStochasticMultistepPolarization2018, khachaturyanMicroscopicInsightsField2024}

Probably the most successful approach to controlling the properties and motion of the DW is by doping and defect engineering.\cite{sunDefectEngineeringPerovskite2021, luIncorporationSpecificDefects, riccaMechanismsPointDefectinduced2022}
On the one hand, defects may act as nucleation centers, lowering the coercive field and increasing the velocities of the walls.\cite{durdievDeterminingThermalActivation2024, mcgillyDynamicsFerroelectric180deg2017}
On the other hand, pinning by defects can slow down the wall dynamics and increase the coercive field for macroscopic switching, so-called hardening.\cite{genenkoMechanismsAgingFatigue2015}
\rv{Also pinching of hysteresis by defects has been reported.}
\cite{saremiLocalControlDefects2018, huangManipulatingFerroelectricBehaviors2021,renLargeElectricfieldinducedStrain2004}
Defects non-commensurate with the lattice have been related to successive pinning and depinning of walls at a disordered energy landscape and thus to jerky or intermittent wall motion.\cite{paruchDomainWallRoughness2005, kimOriginsDomainWall2014, beckerImpedanceSpectroscopyFerroelectrics2022, pramanickDomainsDomainWalls2012, jesseDirectImagingSpatial2008, paruchNanoscaleStudiesFerroelectric2013}
It has been reported that the curvature of the DWs under field-driven motion results from the competition between the local pinning potentials that resist simple sliding and the elastic energy of these propagating interfaces, which tends to maintain flat walls.\cite{ganpulePolarizationRelaxationKinetics2001, tybellDomainWallCreep2002} 
Furthermore, it has been reported that DWs respond mechanically to distortions and fields like a membrane.\cite{placeresjimenezCapacitanceVoltageResponse2013}

\begin{figure*}[tb]
    \centering
    \resizebox{1\linewidth}{!}{
    \begin{Overpic}[abs]{\begin{tabular}{p{.45\textwidth}}\vspace{.21\textheight}\\\end{tabular}}
        \put(0,0){\includegraphics[width=0.45\textwidth]{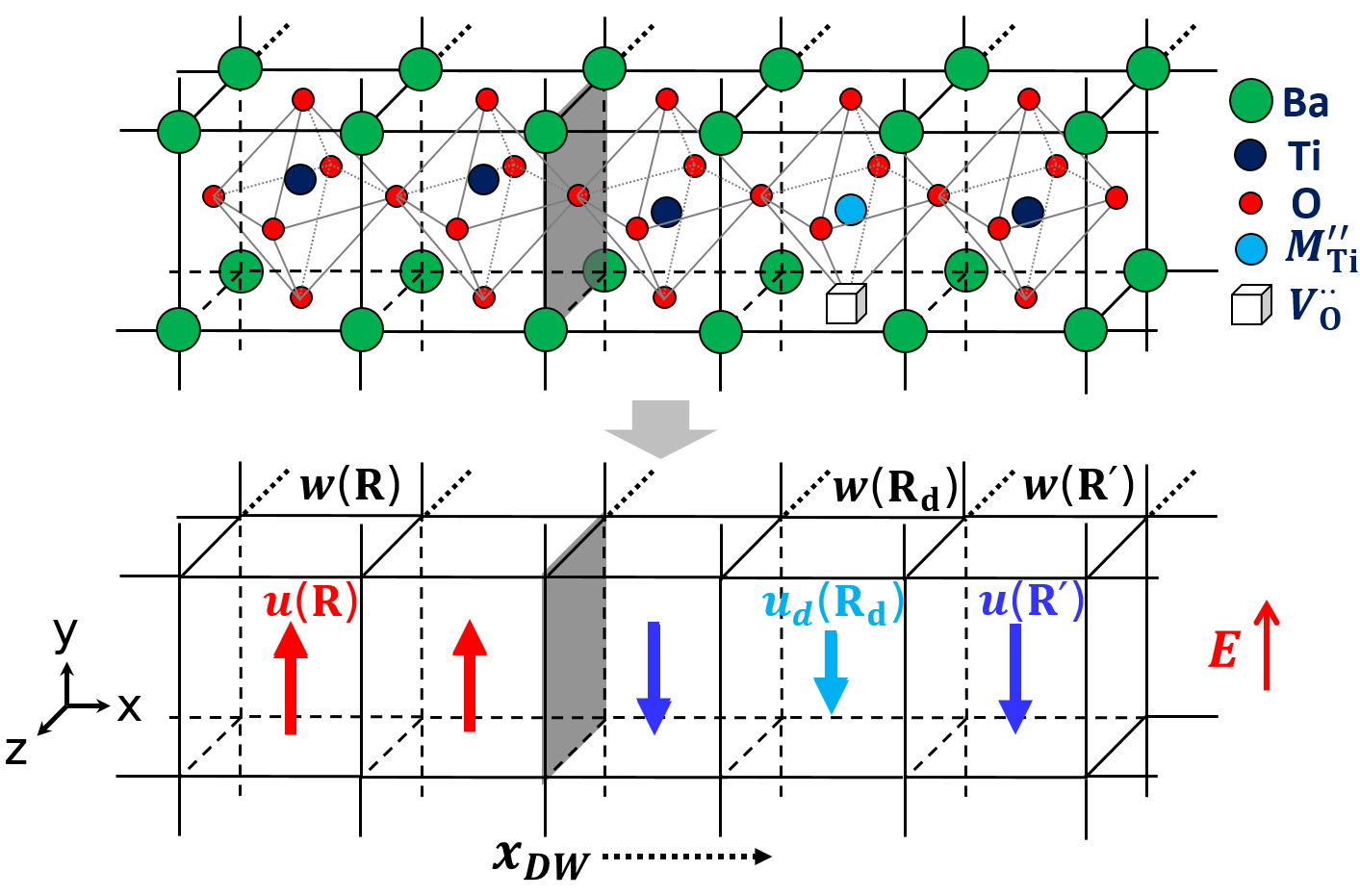}}
        \put(0,145){\textsf{\normalsize(a)}}
        \put(0,65){\textsf{\normalsize(b)}}
    \end{Overpic}
    \begin{Overpic}[abs]{\begin{tabular}{p{.43\textwidth}}\vspace{.21\textheight}\\\end{tabular}}
        \put(0,0){\includegraphics[width=0.43\textwidth]{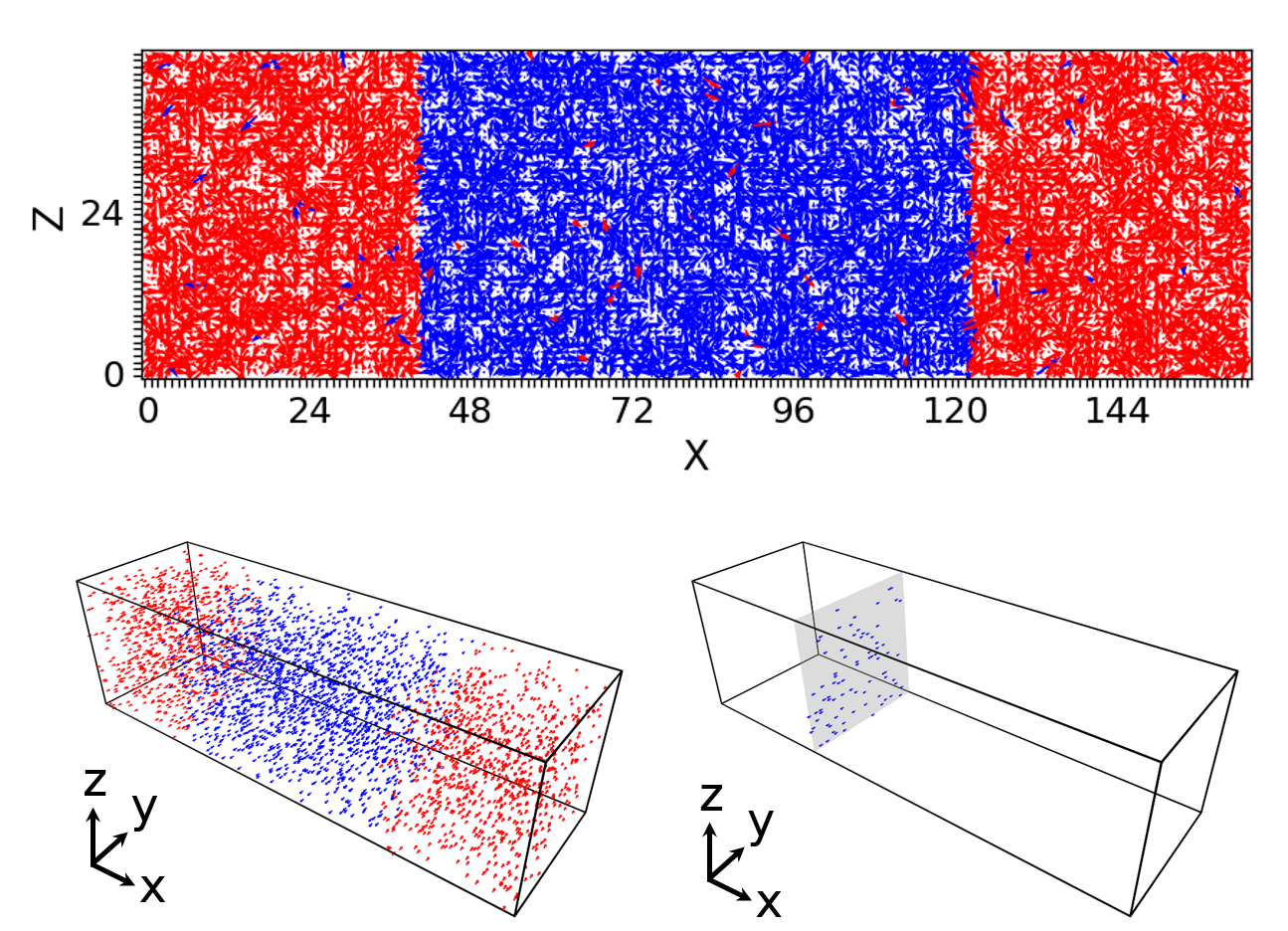}}
        \put(0,145){\textsf{\normalsize(c)}}
        \put(0,65){\textsf{\normalsize(d)}}
        \put(110,65){\textsf{\normalsize(e)}}
    \end{Overpic}
    }
    \caption{(a) Atomic structure of tetragonal BaTiO$_3$ doped with acceptor ions ($M_{\text{Ti}}^{''}$) and oxygen vacancies ($V_{\text{O}}^{\cdot \cdot}$) which form charge-neutral $(M_{\text{Ti}}^{''}$-$V_{\text{O}}^{\cdot \cdot})^{\times}$ defect dipoles. The gray plane indicates a 180$^\circ$ DW.
    (b) In all unit cells \{\textbf{R}\}, all degrees of freedom are mapped on the local strain, $w(\textbf{R})$, and either the free $u(\textbf{R})$, or the frozen-in defect, $u_{d}(\textbf{R})$, dipole moment, as shown by arrows. 
    (c) Cross-cut of the initial tetragonal 180$^\circ$ domain structure at 260~K. 
    (d)--(e) Representative defect distributions for (d) random distribution in space (3D) and (e) defect-rich planes (2D).
    Colors in (b)-(e) encode the sign of the local dipoles $+u_y$ (red) and $-u_y$ (blue).}
    \label{fig:feram_mapping}
\end{figure*}
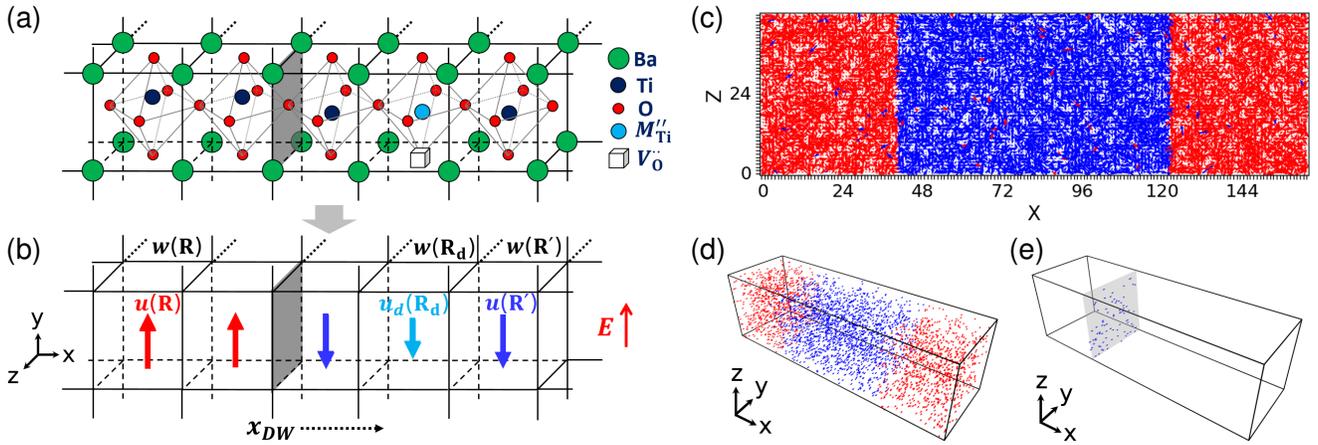

So far, the understanding of the underlying defect-wall coupling is incomplete and various mechanisms have been predicted.
Pinning of walls or their segments has been reported for point defects,\cite{bulanadiInterplayPointExtended2024} vacancies, impurities, dislocations,\cite{zhuoIntrinsicStrainEngineeringDislocation2023} Zr-rich regions in Pb(Zr$_{x}$Ti$_{1-x}$)O$_3$ \cite{paruchDomainWallRoughness2005} or SrTiO$_3$-inclusions in BaTiO$_3$,\cite{dimouPinningDomainWalls2022} and at grain boundaries. \cite{marincelDomainPinningSinglegrain2015, schultheissDomainWallgrainBoundary2020} 
Often not even the types of existing defects in the used samples are known. 
It is furthermore challenging to access nanoscale subsurface defects and domain structures with sufficient time resolution experimentally. 
Instead, atomistic simulations\cite{chandrasekaranDefectOrderingDefectdomainwall2013, liDomainWallMotion2018} and statistical models,\cite{paruchNanoscaleStudiesFerroelectric2013} give access to the fundamental coupling mechanism.
Typically, these distinguish random bond and random field defects.\cite{nattermannInterfacePinningDynamics1990}
Random bond defects, such as substituents with the same valency, can couple with walls via strain, the local modification of the ferroelectric instability, or the energy barrier for switching, but do not break the local symmetry.
For vacancies, large concentrations ($>2$\%) have been studied by density functional theory (DFT) predicting that these pin DWs.\cite{chandrasekaranDefectOrderingDefectdomainwall2013, liDomainWallMotion2018}
% 1/(7*2*3)=2.4%, 1/(6*2*2)=4.2%
In contrast to that, molecular dynamics (MD) simulations at finite temperatures, have revealed that single vacancies can actually promote domain nucleation and switching.\cite{durdievDeterminingThermalActivation2024}
An even larger influence on walls can be expected by random field defects which favor one polarization direction, act as local bias fields, and break the local symmetry.\cite{jesseDirectImagingSpatial2008, genenkoMechanismsAgingFatigue2015}
Surprisingly, their coupling to DWs on the microscopic scale is underrepresented in literature. 
This leads to important gaps in knowledge hindering the nanoscale control of DWs.

In this work we focus on one important subclass of local-field type pinning centers: Charge-neutral acceptor dopant ($M_{\text{Ti}}^{''}$)--oxygen vacancy ($V_\text{O}$) complexes $(M_{\text{Ti}}^{''}$-$V_{\text{O}}^{\cdot \cdot})^{\times}$ in the prototypical ferroelectric material BaTiO$_3$ which are aligned with the surrounding polarization after aging and can only be reoriented through orders of magnitude slower ion migration.
These defects have been reported to act as local restoring forces on the polarization \cite{genenkoMechanismsAgingFatigue2015, renLargeElectricfieldinducedStrain2004} and thereby improving macroscopic functional properties, e.g. reversible dielectric, piezoelectric, and electrocaloric responses.\cite{robelsDomainWallClamping1993, grunebohmInfluenceDefectsFerroelectric2016, liuMultiscaleSimulationsDefect2017, renLargeElectricfieldinducedStrain2004} 
To the best of our knowledge, so far only one MD study on the interaction between charged DWs and defect dipoles in this material has been published.\cite{dhakaneGraphDynamicalNeural2023}
They found that 1~\% oxygen vacancies can slow down the DW motion through the formation of defect dipoles, but these defects have a very localized effect on the surrounding dynamics and may not be able to pin a moving DW due to longer relaxation times.
Here, instead of charged DWs, we \rv{focus on energetically favorable, charge-neutral  DWs. Furthermore, we restrict the study to  walls 
where the ferroelectric polarization rotates by 180$^{\circ}$ while the strain is equal in both adjacent domains, i.e.\ the walls are inelastic or non ferroelastic.} \cite{martonDomainWallsFerroelectric2010}
Using \textit{ab\ initio} derived molecular dynamics simulations, we analyze the field-induced wall motion for different defect distributions in a broad concentration range (0.025 -- 2~\% defects).
We show that the coupling between walls and these defects is short-ranged and thus the defect distribution and particularly the defect-free area in front of the wall is decisive for the wall motion.

%%%%%%%%%%%%%%%%%%%%%%%%%%%%%%

\section{Methods}
\label{sec:Methods}
We perform molecular dynamics simulations using the {\sc{feram}} simulator\cite{nishimatsuFirstprinciplesAccurateTotal2010}, which is based on the effective Hamiltonian approach by Rabe, Zhong and Vanderbilt \cite{zhongPhaseTransitionsBaTiO31994, zhongFirstprinciplesTheoryFerroelectric1995}:

\begin{align}
	H^{\text{eff}}&=
	V^{\text{self}}(\{\bm{u}\})+V^{\text{dpl}}(\{\bm{u}\})+V^{\text{short}}(\{\bm{u}\}) \nonumber\\
	&+V^{\text{elas,homo}}({\eta_1,\eta_2,...,\eta_6})+V^{\text{elas,inho}}(\{\bm{w}\}) \nonumber\\
	&+V^{\text{coup,homo}}(\{\bm{u}\},{\eta_1,\eta_2,...,\eta_6})+V^{\text{coup,inho}}(\{\bm{u}\},\{\bm{w}\})\nonumber\\	
 &+\frac{M_{\text{dip.}}^{*}}{2}\sum_{\bm{R}}\dot{u}_i^2(\bm{R}) 
 -Z^{*}\sum_{\bm{R}}\bm{E}\cdot \bm{u}(\bm{R})\;,
	\label{eq:effham}
\end{align}
where all degrees of freedom are coarse-grained to $\{\bm{u}\}$ and $\{\bm{w}\}$, the soft mode and acoustic displacement vectors in each unit cell (u.c.), respectively, see Fig.~\ref{fig:feram_mapping}~(a)--(b). 
These local soft modes are related to the local polarization $\bm{p}$ as $\bm{p}=Z^{*}\bm{u(\bm{R})}/\Omega$, with $Z^{*}$ and $\Omega$ being the effective Born charge of the unit cell and its volume.
The potential energy terms are $V^{\text{self}}(\{\bm{u}\})$, the self-energy of the local soft modes, $V^{\text{dpl}}(\{\bm{u}\})$ and $V^{\text{short}}(\{\bm{u}\})$, their long-range and short-range interactions, $V^{\text{elas,homo}}(\eta_1,\eta_2,...,\eta_6)$ and $V^{\text{elas,inho}}(\{\bm{w}\})$, the elastic energies from homogeneous and inhomogeneous strain, with $\eta_{1},..., \eta_{6}$ being the six components of the homogeneous strain tensor in Voigt notation ($\eta_1=e_{xx}$, $\eta_4=e_{yz}$), $V^{\text{coup,homo}}(\{\bm{u}\},\eta_1,\eta_2,...,\eta_6)$ and $V^{\text{coup,inho}}(\{\bm{u}\},\{\bm{w}\})$, which is the coupling between the local soft modes and homogeneous and inhomogeneous strain. 
The last term couples the local soft modes to an external electric field.
As homogenous strain and $\{\bm{w}\}$  are internally optimized, only the kinetic energy of the local soft modes, with an effective mass $M^{*}_{\text{dip.}}$, is explicitly included in the simulations.
Together with the \textit{ab\ initio} parameterization from Ref.~\onlinecite{nishimatsuFirstprinciplesAccurateTotal2010}, this model can successfully describe the full temperature-phase diagram of BaTiO$_3$. 
However, it underestimates experimental transition temperatures and the tetragonal phase is (meta-)stable between 180~K and 275~K (140~K -- 295~K).
This has been related to the underlying DFT data,\cite{nishimatsuFirstprinciplesAccurateTotal2010} missing anharmonicities (e.g., thermal expansion), and missing higher-energy phonon modes. \cite{mayerFinitetemperatureInvestigationHomovalent2022, mayerHiddenPhasesHomovalent2023}

Without loss of generality, we align the tetragonal axes along $y$ and thus set the components of the time-averaged local polarization to $\langle p_y\rangle\neq0$ and $\langle p_x\rangle=\langle p_z\rangle=0$.
We then approximate $(M_{Ti}^{''}$-V$_O^{\cdot \cdot})^{\times}$ complexes by frozen-in defect dipoles ($u_d$)\rv{, as these defect dipoles at ambient temperatures have orders of magnitude  slower dynamics than the free dipoles or domain wall.} \cite{renLargeElectricfieldinducedStrain2004, erhartFormationSwitchingDefect2013, genenkoMechanismsAgingFatigue2015}. 
The strength of these dipoles is adjusted by supplementary DFT calculations for the  transition metal ions $M$:
Cu$^{+2}$, Mn$^{+2}$, and Fe$^{+2}$ using the VASP code \cite{kresseEfficientIterativeSchemes1996} together with PBEsol potentials and Hubbard $U$ corrections of 6.52, 7.2 and 6.8~eV, respectively. 
The used $2\times 2 \times 2$ supercells with one defect dipole each correspond to a defect concentration of approximately 4~\%. 
To reduce finite-size effects, the simulation cell is constrained to the tetragonal structure of pristine BaTiO$_3$, while atomic positions are relaxed with a force threshold of 0.01 eV/\AA{} using a plane-wave cutoff energy of 520~eV and \textit{k}-point sampling of $2\times2\times2$. 
Using this approach, we find defect strengths of 29.6 (24.9), 29.1 (25.1), and 29.5 (25.3)~$\mu$C/cm$^2$ for Cu$^{+2}$, Mn$^{+2}$, and Fe$^{+2}$ for parallel (antiparallel) alignment of defect dipoles and surrounding polarization $P_y$ and 30.9~$\mu$C/cm$^2$ for pristine BaTiO$_3$. 
Thus, antiparallel defects \rv{reduce} the local polarization roughly by a factor of 0.95 (0.81).
For simplicity, we fix the defect dipole to 24.7~$\mu$C/cm$^2$, i.e.\ which is a factor of 0.8 smaller than $P_y$ in the ferroelectric phase.\footnote{Note that the DFT informed defect strength is an approximation which does not include the change of this ratio with temperature \rv{nor with the direction of the surrounding polarizations}. Close to the transition temperature, e.g.\ at 280~K the ratio of the chosen defect strength and the polarization of the pristine material increases to 0.9. \rv{Furthermore, we neglect the possible strain field and changes of strain--polarization coupling around the defects. Both approximations could change the predicted domain wall velocities and pinning concentrations quantitatively.}}
As illustrated in Fig.~\ref{fig:feram_mapping}~(d)--(e), we only consider defect dipoles initially aligned with the polarization direction of the surrounding domains, i.e.\ along $+y$ in $+p_y$ domains or along $-y$ in $-p_y$ domains. 
%both for randomly distributed defect dipoles (0.5, 1, and 2~\%) and defect-rich planes (with 1, 2, 2.5, 3, 3.5, 4, 5, 6, and 7~\% defects).
%\AG{@Aris: Thank you} 

For our large-scale MD simulations, we use a system size of $164 \times 48 \times 48$~u.c.\ of BaTiO$_3$ together with periodic boundary conditions and the Nos\'{e}-Poincar\'{e} thermostat with a timestep of 1~fs. 
If not stated otherwise, we discuss results for DWs at 260~K.
We initialize equidistant domains with $+p_y$ and $-p_y$, separated by tetragonal 180$^\circ$ DWs normal to $x$, see Fig.~\ref{fig:feram_mapping}, by applying \rv{auxiliary} fields $E_y=\pm100$~kV/cm for 30~ps. 
\footnote{\rv{One can initialize domain structures in different ways, e.g.\ by introducing nuclei, start to equilibrate from ideal domain structure at 0~K (cf.\ $Domain.py$ in the Github repository, \url{https://github.com/kazats/AutoFeram}), or here, inspired by piezoresponse force microscopy, by applying local auxiliary fields to write in the domain structure. After further equilibration, the properties of the system do not depend on the chosen initialization, e.g.\ reducing the auxiliary fields by one order of magnitude does not modifiy the domain structure.}}
\rv{This initial state is equilibrated without a field for 30~ps to reach thermal equilibrium. Afterwards, }
we instantaneously apply a \rv{homogeneous} field $E_y=\rv{+}100$~kV/cm for 100~ps.
This field strength is sufficient to depin DWs from the lattice in the temperature range of interest (240--280~K).
Note that it is expected that experimental field strengths are overestimated in simulations on idealized materials and without explicit treatment of the electrodes.\cite{jiangResolvingLandauerParadox2009, gaoRevealingRoleDefects2011}

In order to characterize the DWs, we identify the center of the walls ($x_0$) with u.c.\ resolution by the sign change of $p_y$, the $y$-component of the local polarization, after taking moving averages over unit cells along the $x$-direction.
This technique is similar to box-blur in image processing.
As proven for one example, the results are converged with respect to the chosen convolution kernel size of 11~u.c.
The mean square wall roughness $R_{DW}$ is given as 
\begin{equation}
R_{\text{DW}}=\sqrt{\sum_i^N \frac{(x_0(i)-X_{\text{DW}})^2}{N}},
\end{equation}
with the sum running over all $N=48\times 48$~u.c.\  segments on $x$-planes and  $X_{\text{DW}}=\langle x_0\rangle$ being the mean value of $x_0$.
As the wall roughness is not a measure for the wall width, we supplementary determine  the mean widths ($d_{\text{DW}}$) of the walls based on the mean polarization per $x$-layer $P(x)$ with 
\begin{equation}
P(x)=P_y \tanh \left(\frac{x-X_0}{2d_{\text{DW}}}\right), \cite{foethComparisonHREMWeak1999}
\end{equation}
where $P_y$, $X_0$, and $d_{\text{DW}}$ are fit parameters. 
Note that this equation can only estimate the wall width for flat walls, while $d_{\text{DW}}$ is overestimated in case of wall bending.

To estimate critical dilations of nuclei on the moving wall, we furthermore analyze the time evolution of dipole clusters switching from $-p_z$ to $+p_z$ on separate $x$-planes (slices) in 9 independent simulations in the time interval $3-60\,$ps. 
\begin{figure*}[t]
    \centering
    \resizebox{.9\linewidth}{!}{
    \begin{Overpic}[abs]{\begin{tabular}{p{.23\textwidth}}\vspace{.18\textheight}\\\end{tabular}}
    	\put(0,0){\includegraphics[height=0.177\textheight]{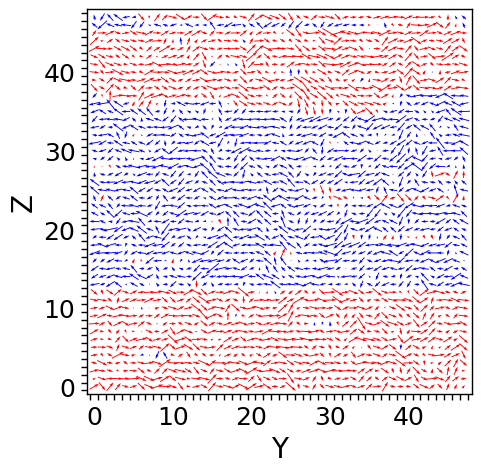}}
    	\put(10,125){\textsf{(a)}}
    \end{Overpic}
    \begin{Overpic}[abs]{\begin{tabular}{p{.135\textwidth}}\vspace{.18\textheight}\\\end{tabular}}
    	\put(0,0){\includegraphics[height=0.2\textheight]{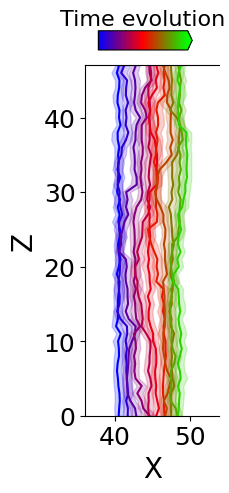}}
    	\put(0,125){\textsf{(b)}}
    \end{Overpic}
    \begin{Overpic}[abs]{\begin{tabular}{p{.18\textwidth}}\vspace{.18\textheight}\\\end{tabular}}
    	\put(2,7){\includegraphics[height=0.17\textheight,clip=true,trim=0cm 0cm 0cm 0cm]{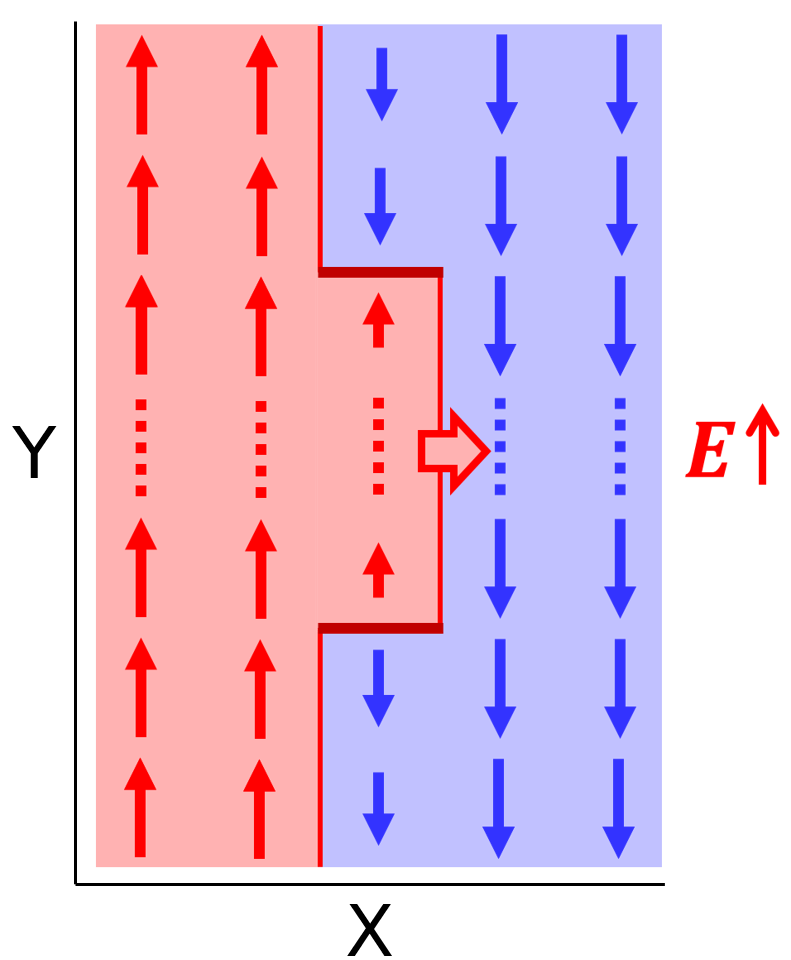}}
    	\put(0,125){\textsf{(c)}}
    \end{Overpic}
    \begin{Overpic}[abs]{\begin{tabular}{p{.18\textwidth}}\vspace{.18\textheight}\\\end{tabular}}
    	\put(2,7){\includegraphics[height=0.17\textheight,clip=true,trim=0cm 0cm 0cm 0cm]{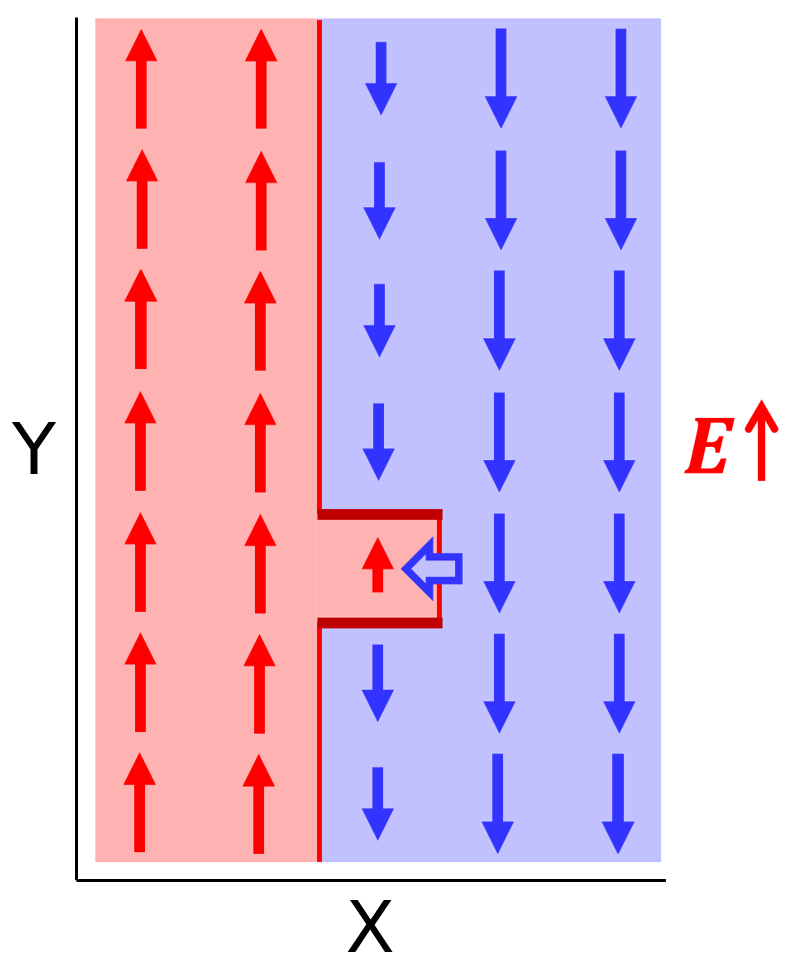}}
    	\put(0,125){\textsf{(d)}}
    \end{Overpic}
    }
    \caption{DW motion in the pristine material at 240~K in an external field of $E_y=100$~kV/cm.
    (a) Snapshot of the DW interface (layer $x=41$) 10~ps after field application hosting 2D stripe domains.
    (b) Change of the local DW positions averaged over $y$ ($x_{0}(z)$) with time.
    Colors encode the time from 0~ps (blue) to 100~ps (green) in intervals of 10~ps. 
    (c)--(d) Schematic illustration of domain growth along $x$ (wide arrow) (c) and dipole fluctuations and disappearing clusters below the critical size of a nucleus (d) on a $z$-plane.
    Blue and red arrows depict $-p_y$ and $+p_y$ dipoles.}
    \label{fig:dw_pt_240K}
\end{figure*}
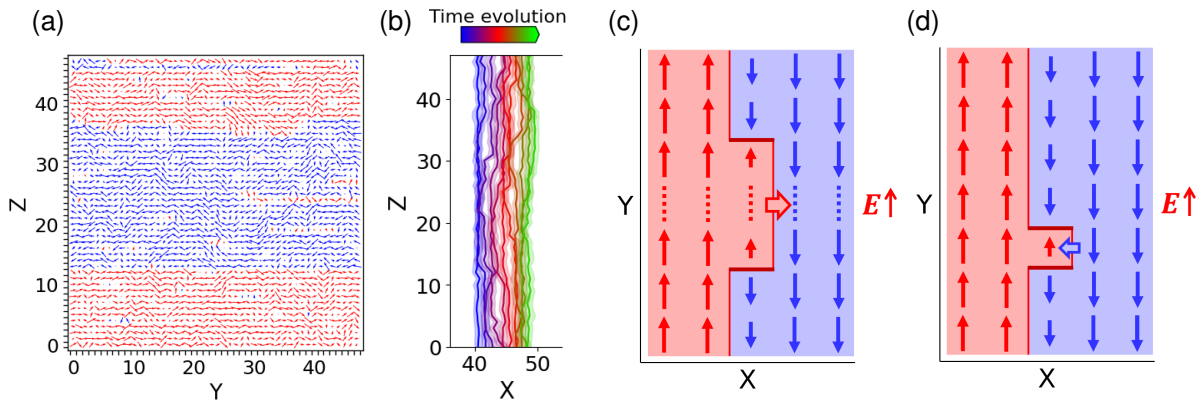
%\BU{%The time interval is 3 to 60 ps.
%The data column I received from Majid ranges from 1,000 to 60,000. I assume this is in femtoseconds. %Additionally, 9 independent simulations were considered.
%}
We define a cluster as a connected region of unit cells on one plane and record statistics over the evolution of clusters on different planes. 
Cluster analysis is performed through a wrapper around the labeling capabilities implemented in \texttt{skicit-image}.~\cite{waltScikitimageImageProcessing2014}
To distinguish thermal fluctuations and nucleating clusters, we filter out all clusters which exist for only one timestep (1~ps).
We then extract the critical in-plane dilations of nuclei, and their standard deviations.  
Details of the implementation and a description of procedure including outlier detection can be found in Refs.~\onlinecite{tengICAMSSFCOrtho_2023GitLab2023, khachaturyanMicroscopicInsightsField2024}. 
The only change with respect to the provided reference is an adaption to tetragonal symmetry.

%%%%%%%%%%%%%%%%%%%%%%%%%%%%%%

\section{Results \& Discussion}
\label{sec:Results}

\subsection{Domain wall motion in the pristine material}
\label{ssec:pristine}

To understand how defects modify DWs and their motion, the field-response of the pristine material (no defects) is needed as reference.
Without an applied field, the mean DW roughness (width) between 240~K and 280~K is 0.8 -- 1.2~u.c.  (0.6 -- 1~u.c.), i.e.\ less than 0.5~nm, in agreement with predictions from literature. \cite{martonDomainWallsFerroelectric2010, grunebohmDomainStructureTetragonal2012}
There is no driving force for wall motion and temporary broadening or shifting of walls are rare events for the given temperature range.\cite{shinNucleationGrowthMechanism2007, klompThermalStabilityNanoscale2022}
An external field $E_y$, parallel to the $+p_y$ domain, acts as a driving force for the growth of this domain.
As discussed in Refs.~\onlinecite{bodduMolecularDynamicsStudy2017, khachaturyanDomainWallAcceleration2022}, one can distinguish three field regimes: 
(i) Pinning by the lattice: The energy barrier to shift the walls from BaO planes to TiO$_2$ planes is large compared to the activation energy related to the applied field. 
The motion of the wall is a rare event and the macroscopic ferroelectric polarization does not change with time (no switching). 
(ii) Switching by thermally activated DW propagation along $x$-direction is activated by $E_y \geq 50$~kV/cm. 
With increasing field strength, the velocity of the walls and non-equilibrium effects after the fast field application, increase.
The latter induces temporary sign change of dipoles in the blue domain and thereby the boost of the initial wall velocity. 
This is the precursor for regime (iii), the additional homogeneous nucleation in the antiparallel domain and fast switching by 3D growth of these nuclei which appear for more than about 400~kV/cm.

In the remainder of the paper we focus on field-regime (ii), the thermally activated wall motion in $E_y=100$~kV/cm. 
In agreement to predictions by atomistic and coarse-grained simulations in Refs.~\onlinecite{klompThermalStabilityNanoscale2022, shinNucleationGrowthMechanism2007}, this motion starts with enhanced dipole fluctuations on the wall and the nucleation of 2D clusters of dipoles aligned with the field. 
Our cluster analysis yields lower bounds for the critical size of these nuclei of about $12.3\pm 1.3$ and $4.2\pm 0.2$ u.c.\ along the polarization direction ($y$) and perpendicular to it ($z$).
% pre-rounding results: y\_ext\_c1 = $4.24\pm 0.21$ and z\_ext\_c1 = $12.3\pm 1.31$
After rapid expansion along $y$, 2D stripe domains are formed on the walls, see Fig.~\ref{fig:dw_pt_240K}~(a), which grow and merge by slower sideway motion along $z$ until the whole plane has switched. 
With time, the walls move through the material by this thermally activated process, see Fig.~\ref{fig:dw_pt_240K}~(b).
Thereby, the probability to nucleate a 2D cluster in front of the walls is constant on all $x$-planes and in different simulation runs. 
The $x$-interface is thus no longer flat, different segments of the wall have propagated more than others and the positions of the wall center ($x_0(z)$) deviate up to 3.4~u.c, \rv{see} Fig.~\ref{fig:dw_pt_240K}~(b).
%The maximal variations max(xDW)-min(xDW) of wall center along $z$ reach  2.8, 3.4 and 4.2~u.c.\ for 240~K, 260~K and 280~K. 

The anisotropies of the critical cluster dilations and the cluster growth are related to the large additional energy penalty to form charged interfaces (depolarization) along the polarization direction $y$. 
On the one hand, interfaces along $x$ and $z$ are charge-neutral and low in energy. 
On the other hand, interfaces along $y$ are charged and the depolarization fields induce large energy penalties, see Fig.~\ref{fig:dw_pt_240K}~(c)--(d).\cite{khachaturyanDomainWallAcceleration2022, klompThermalStabilityNanoscale2022}

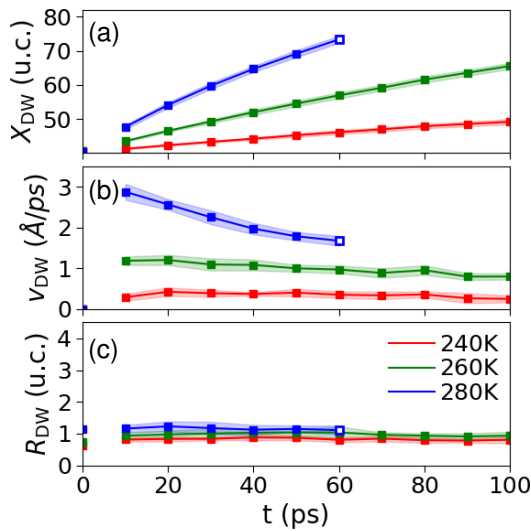
\begin{figure}[t]
    \centering
    %\resizebox{.9\linewidth}{!}{
    \begin{Overpic}[abs]{\begin{tabular}{p{.4\textwidth}}\vspace{.29\textheight}\\\end{tabular}}
        \put(0,0){\includegraphics[width=0.4\textwidth]{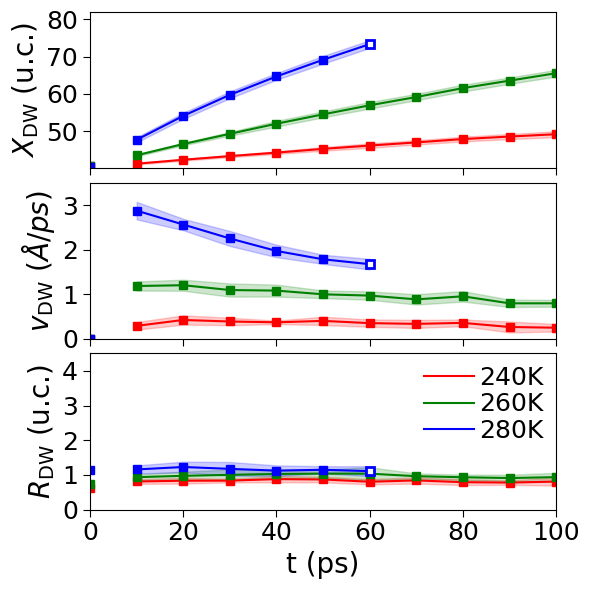}}
        \put(33,188){\textsf{\normalsize(a)}}
        \put(33,128){\textsf{\normalsize(b)}}
        \put(33,68){\textsf{\normalsize(c)}}
    \end{Overpic}
    %}
    \caption{Temperature dependence of DW (a) position, $X_{\text{DW}}$, (b) velocity, $v_{\text{DW}}$, and (c) roughness, $R_{\text{DW}}$, of pristine BaTiO$_3$ at 240~K (red), 260~K (green) and 280~K (blue).
    At each temperature, the mean values (solid line) and standard deviation (shaded area) of 10 independent simulations are shown.
    At 280~K, both DWs get too close for further analysis after 60~ps or 70~ps (empty squares).}
    \label{fig:xvr_pristine}
\end{figure}

In the steady-state, the velocity of the wall motion increases from 0.4~\AA/ps at 240~K by factors of approximately 2.5 and 4 if going to 260~K and 280~K, as shown in Fig.~\ref{fig:xvr_pristine}~(b).\footnote{Note that the polarization in the $+p_y$ domain and the wall velocity increase and decrease with time. While the change of $p_y$ is negligible at low temperatures, it reaches 5~\% and 14~\% at 260~K and 280~K.}
% 260K: (1.0~\AA/ps), 280K: (1.7~\AA/ps) 
Note that at 280~K we can only monitor the walls for 60~ps. 
Afterwards they have crossed the simulation cell and annihilated the $-p_y$ domain.
For this temperature, we furthermore observe a non-equilibrium boost of the initial velocity already for this moderate field strength as initially about 3.5~\% dipoles in the center of the negative domain are aligned with the field direction.  
With time these dipoles switch back until the steady state (about 2.4~\% $+p_y$ in the negative domain) is reached at around 40~ps. 
The fact that we find non-equilibrium switching at the moderate field strength is related to enhanced thermal fluctuations and reduced energy barriers for polarization switching close to the cubic to tetragonal phase transition of our model. 
Well in the tetragonal phase \rv{i.e.\ at 260 or 240~K thermal fluctuations are not sufficient to cross the barrier for initial switching for the moderate field strength, cf.\ Fig.~\ref{fig:uhist_260_280K}}, and the steady state of the wall motion is already reached within the first 10~ps.
The mean roughness of the walls is only about 1~u.c.\ and does not change significantly either with temperature or time, or between independent samples, see Fig.~\ref{fig:xvr_pristine}~(c).

In summary, DW motion in the pristine material follows the nucleation and growth of 2D clusters, which form at all points of the DW with equal probability.
The velocity of this process is sensitive to temperature and field strength, while the DW roughness is less sensitive to these parameters.

\subsection{Defect dipoles and wall motion: Random distribution}
\label{ssec:r3d}
How do randomly distributed defect dipoles aligned with the initial domain structure, see Fig.~\ref{fig:feram_mapping}~(d), affect the wall motion?
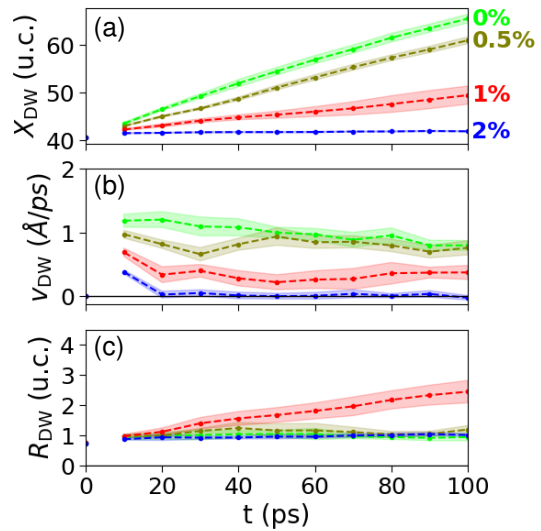
\begin{figure}[t]
    \centering
    %\resizebox{.9\linewidth}{!}{
    \begin{Overpic}[abs]{\begin{tabular}{p{.4\textwidth}}\vspace{.29\textheight}\\\end{tabular}}
        \put(0,0){\includegraphics[width=0.4\textwidth]{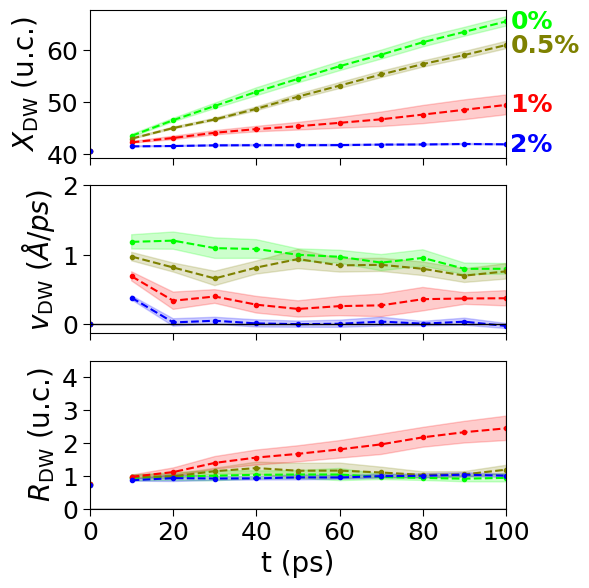}}
        \put(34,190){\textsf{\normalsize(a)}}
        \put(34,130){\textsf{\normalsize(b)}}
        \put(34,69){\textsf{\normalsize(c)}}
    \end{Overpic}
    %}
    \caption{Concentration dependence of DW (a) position, $X_{\text{DW}}$, (b) velocity, $v_{\text{DW}}$, and (c) roughness, $R_{\text{DW}}$, for 0~\% -- 2~\% (colors) randomly distributed frozen-in defect dipoles aligned with the initial polarization direction at 260~K.
    Mean values (solid lines) and standard deviations (shaded areas) of 10 independent simulations are shown.}
    \label{fig:xdw_r3d_conc}
\end{figure}

As defect dipoles do not switch by thermal fluctuations or the applied field, one may expect that they induce local bias fields and pin the walls. 
To test this hypothesis, we analyze the wall properties for different defect concentrations, see Fig.~\ref{fig:xdw_r3d_conc}.
%\AG{1\% P0 descreases by about 1\% 2\% decreas by about 2\% thus no macroscopic increase of P; }
%\SH{20241219: at t=0ps, mean P0 at parallel regions (x=0:30) for 0.5\%: decreases by 0.7\%, 1\%: by 1.2\%, 2\%: by 2.1}
Indeed the mean wall velocity decreases systematically with defect concentration, i.e.\ by about 5~\% and 50~\% for 0.5~\% and 1~\% defects, and the critical density for full pinning is below 2~\% defects, see Fig.~\ref{fig:xdw_r3d_conc}~(b). 
Furthermore, already 0.5~\% defects suppress the initial boost of the DW velocity at 280~K, cf.\ appendix Fig.~\ref{fig:xvr_r3d.5p}.
%\AG{here we need a comment that at 10 ps the domain wall only interacts with few defect planes and thus the velocity reduction is still smaller, cf. next section  }
\rv{Note that the DW only interacts with few defect planes before 10~ps, and thus the velocity reduction is still smaller for all defect concentrations, c.f.\ Sec.~\ref{ssec:1L}.}
For a given defect concentration, the wall velocity furthermore decreases with the strength of the defects,  e.g.\ to $<0.1$~\AA/ps for $|u_d|=0.15$~\rv{\AA} ($1.33\rv{\cdot}P_y$) and 1\% defects, cf.\ appendix Fig.~\ref{fig:xvr_r3d1p_ud}, and a smaller critical concentration for pinning can be expected. 

\begin{figure}[t]
    \centering
    \begin{Overpic}[abs]{\begin{tabular}{p{.33\textwidth}}\vspace{.24\textheight}\\\end{tabular}}
        \put(0,0){\includegraphics[height=0.25\textheight]{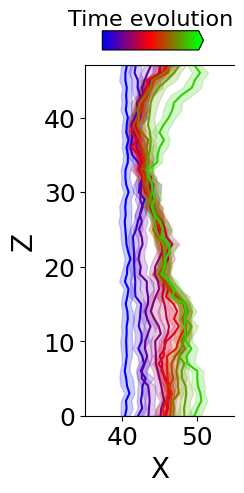}}
        \put(0,160){\textsf{\normalsize(a)}}
        \put(86,0){\includegraphics[height=0.25\textheight]{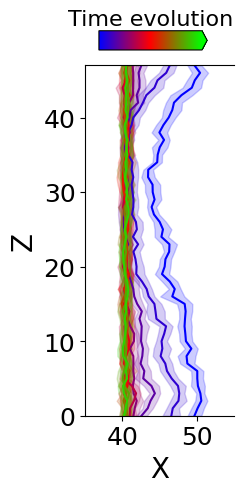}}
        \put(158,40){\color{black}\vector(-1,0){15}}
        \put(151,125){\color{black}\vector(-1,1){15}}
        \put(144,105){\color{red}\large$\boldsymbol{E}$=0}
        \put(86,160){\textsf{\normalsize(b)}}
    \end{Overpic}
    \caption{(a) Change of the DW profile $x_{0}(z)$ with time in the presence of randomly 1~\% distributed defects at 260~K (a) in an external field along $y$ and (b) after the field has been removed again. 
    Colors encode the time from 0~ps (blue) to 100~ps (green) in intervals of 10~ps after field removal and the first configuration in (b) corresponds to the final one in (a).}
    \label{fig:dw_r3d1p}
\end{figure}
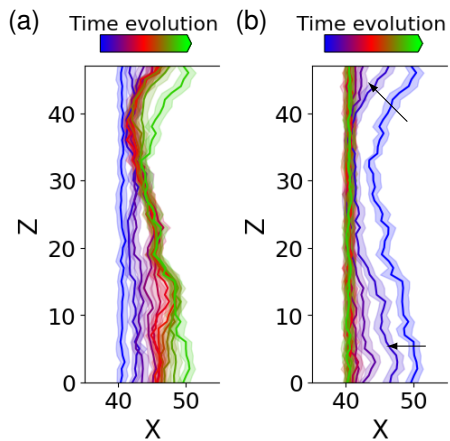

Not only the DWs but also the polarization in the domains is modified by the defect dipoles.
For the DFT-informed defect strength corresponding to $0.8 \cdot P_y$, the mean polarization in both domains is reduced by 0.7~\%, 1.2~\% and 2.1~\% for 0.5, 1 and 2~\% defects. 
If defects and free dipoles are locally antiparallel to each other after wall motion, the local polarization is reduced further, e.g.\ by 13~\% for 1~\% defects.
Stronger defects, e.g.\ $|u_d|$=0.15~\AA{} corresponding to $1.33\rv{\cdot}P_y$, enhance the initial polarization by about 1~\%.
%\SH{20241219: at 100ps, mean P0 in antiparallel regions (x=41:58) for 0.5\%: decreases by 7\%, (x=41:43) 1\%: by 13\%, 2\%: pinned}
%\SH{20250109: at 0ps, mean P0 for $|ud|$=0.11A: increases by 1\%, $|ud|$=0.15A: increases by 1\%; at 100ps: no completely switched in antiparallel domain, almost pinned}
Macroscopically, the defects distributed in 3D space thus indeed act on the system as bias fields and one may expect an enhanced coercive field experimentally. 
As the critical density for full pinning of 2~\% is the lower bound of the defect concentration used in many atomistic studies, these predict wall pinning by acceptor dopants. 
For slightly smaller defect concentrations, however, already a moderate field is sufficient to depin the walls, cf.\ Fig.~\ref{fig:dw_r3d1p}~(a). 
Fully in line with the interpretation of defect-induced bias fields, 1~\% randomly distributed defects also restore the initial wall position and shape once the external field is removed, see Fig.~\ref{fig:dw_r3d1p}. 
Even though some wall segments have moved by more than 10~u.c. in the field (a), the wall relaxes back to its initial position in about 40~ps (b).
Our simulations can thus prove the predictions in Ref.~\onlinecite{renLargeElectricfieldinducedStrain2004} that point defects may act as a restoring force on the domain structure and are thus beneficial for large reversible functional responses.

In addition to these macroscopic changes, the defects also induce time-dependent variations of wall velocities and roughening, see Fig.~\ref{fig:xdw_r3d_conc}~(b)--(c). 
For the intermediate concentration of 1~\%, the mean wall roughness even increases systematically with time and after 100~ps it is twice the value of the pristine material.
Furthermore, wall velocity and roughening vary between statistically independent simulations (colored backgrounds) with maximal deviations of up to 47~\% and 15~\%, see shaded areas in Fig.~\ref{fig:xdw_r3d_conc}.
% 47~\% %(0.17~\AA/ps), 15~\% %(0.37~u.c.) 
Even larger deviations have to be expected between different random defect distributions, particularly if of different homogeneity. 

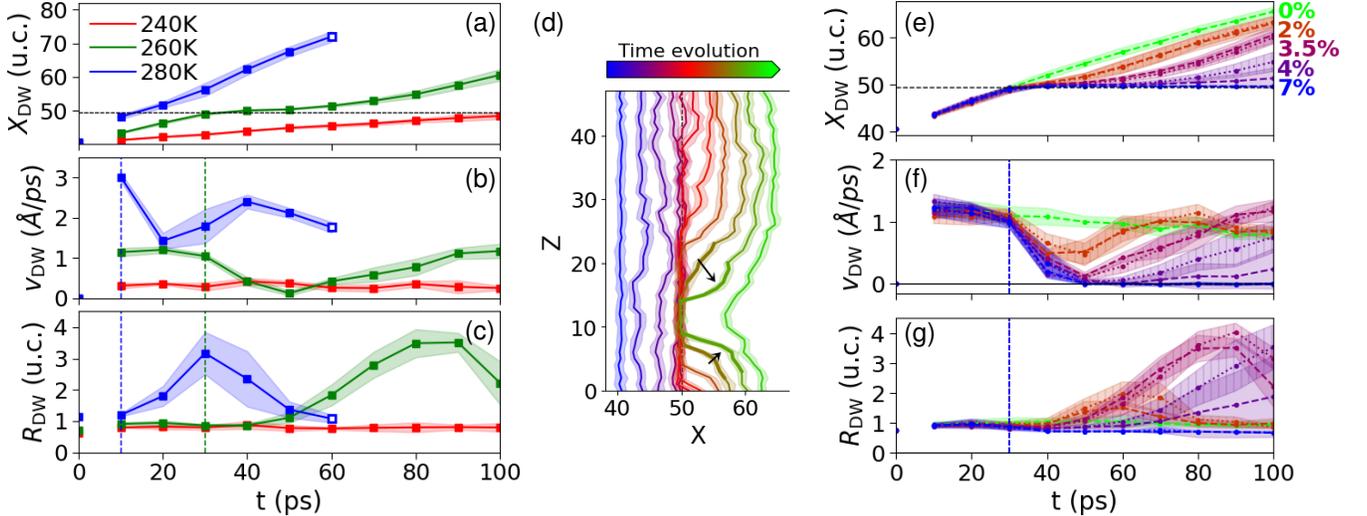
\begin{figure*}[t]
    \centering
    \resizebox{\linewidth}{!}{
    \begin{Overpic}[abs]{\begin{tabular}{p{.39\textwidth}}\vspace{.29\textheight}\\\end{tabular}}
        \put(0,0){\includegraphics[width=0.4\textwidth]{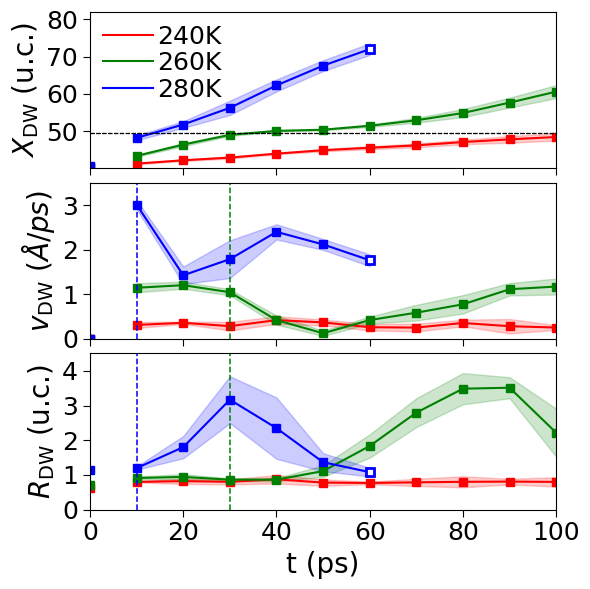}}
        \put(179,189){\textsf{\normalsize(a)}}
        \put(179,129){\textsf{\normalsize(b)}}
        \put(179,71){\textsf{\normalsize(c)}}
    \end{Overpic}
    \begin{Overpic}[abs]{\begin{tabular}{p{.2\textwidth}}\vspace{.29\textheight}\\\end{tabular}}
        \put(0,26){\includegraphics[height=0.24\textheight]{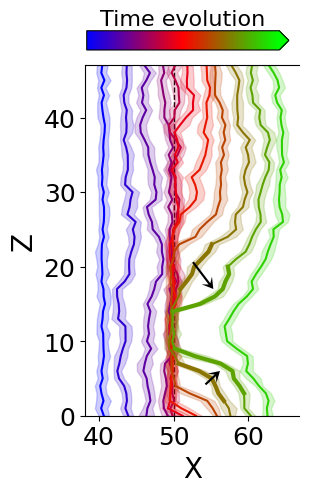}}
        \put(0,189){\textsf{\normalsize(d)}}
    \end{Overpic}
    \begin{Overpic}[abs]{\begin{tabular}{p{.38\textwidth}}\vspace{.29\textheight}\\\end{tabular}}
        \put(0,0){\includegraphics[width=0.4\textwidth]{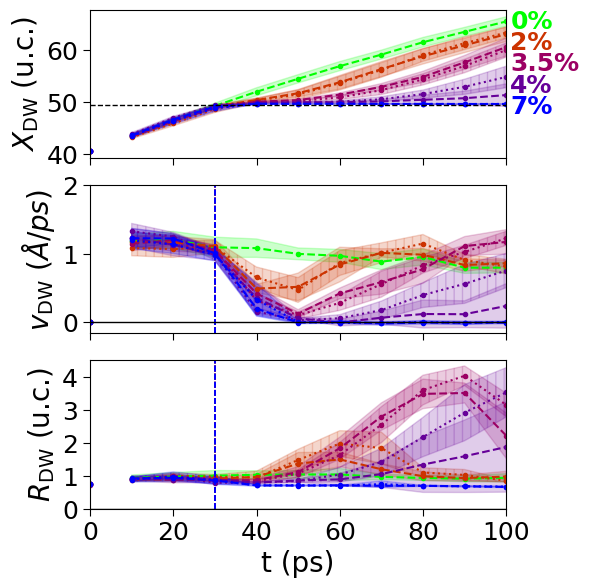}}
        \put(34,189){\textsf{\normalsize(e)}}
        \put(34,129){\textsf{\normalsize(f)}}
        \put(34,70){\textsf{\normalsize(g)}}
    \end{Overpic}
    }
    \caption{DW (a)/(e) position, $X_{\text{DW}}$, (b)/(f) velocity, $v_{\text{DW}}$, and (c)/(g) roughness, $R_{\text{DW}}$, in the presence of defect dipoles ($u_d=-0.09$~\AA) randomly distributed on plane $x=50$~u.c., cf.\ Fig.~\ref{fig:feram_mapping}~(e), for (a)--(c) 3.5~\% defects and varying temperature and (e)--(g) at 260~K and varying defect concentration.
    For (a)--(c) or (e)--(g), mean values (solid line) and standard deviation (shaded area) of 1 selected defect distribution or 2 defect distributions (dotted and dashed lines) with 5 independent simulations each are shown.
    At 280~K both DWs get too close for further analysis after 70~ps (empty square). 
    The vertical lines in (b)--(c) mark the times when the center of the DWs reaches the defect plane.
    (d) Corresponding changes of the DW profile $x_{0}(z)$ with time for 3.5~\% at 260~K.
    Colors encode the time after field application from 0~ps (blue) to 100~ps (green) in intervals of 10~ps.}
    \label{fig:xvr_1L3.5p}
\end{figure*}
These observations are a consequence of local pinning and sequential depinning of wall segments at defect-rich regions. 
For low defect concentrations, the defect configuration in most layers only interacts weakly with the moving wall. 
Particularly there is no tendency for wall pinning at separate defects. 
Instead, the wall segments flow around these defects.
The main influence of 0.5~\% defects is thus to reduce the local probabilities for nucleation and growth of new clusters and the forward motion of parts of the wall. 
A random distribution of 1~\% defects, however, temporarily pins a significant fraction of wall segments. 
As the pinned segments for 1~\% defects are at different positions on different $x$-layers the roughening increases with time and is related to the bending of the wall, see Fig.~\ref{fig:dw_r3d1p}~(a). 
The velocity of the wall reaches a steady state as soon as the fraction of newly pinned and successively depinned segments balances for a sufficiently homogeneous distribution of defects. 
Due to variations of the local defect density, also the wall in the material with 0.5~\% defects reaches a plane with 1~\% defects at about 25~ps (see appendix Fig.~\ref{fig:defects_r3d.5p_x})\rv{,} where a large fraction of segments is temporarily pinned. 
The mean wall velocity shows a local minimum and the wall roughness is enhanced for about 20~ps, see Fig.~\ref{fig:xdw_r3d_conc}.
With time, the whole wall is depinned from the corresponding plane and initial wall roughness and velocity are restored while the wall passes layers with lower defect concentrations.  
\rv{As shown in appendix Fig.~\ref{fig:defects_r3d.5p_x}, the distance traveled by a domain wall through regions without such local enhanced defect density is larger. 
It should be noted that there is however no one-to-one correspondence between local defect density and local velocity. Instead the velocity depends on the local concentration on the plane and the planes passed before, as well as on the distribution of defects on the planes as discussed in the next section.} 
In case of 2~\% defects, some segments can penetrate defect-poor regions in the first 10~ps. 
The average wall position moves by 1~u.c and the roughness increases by about 18~\%.
Afterwards the local defect density and the defect distribution in $x$-planes do not allow any further penetration of wall segments. 

In summary, homogeneously distributed defect dipoles indeed induce internal bias fields,  which act on the macroscopic field response as if the applied field had been reduced\rv{,} and which restore the initial domain structure. 
On the nanoscale, however, it depends on the local density and distribution of defects which wall segments are pinned and whether this pinning can persist with wall roughening and slowing down of the wall motion, or whether the walls can flow around separate defects.

\subsection{Domain wall motion and defect plane}
\label{ssec:1L}
To gain a full microscopic understanding of the underlying local pinning and depinning processes, we turn to a defect distribution with different symmetry: 
A defect-rich plane (2D defect distribution) with $-p_y$ defects is added in front of the wall, see Fig~\ref{fig:feram_mapping}~(e).
We answer the following questions: How long-ranged is the interaction between walls and defects? 
How do the defects pin and bend the wall? 
What is the local critical defect density to pin a wall? 

Wall properties for different temperatures are compared in Fig.~\ref{fig:xvr_1L3.5p}~(a)--(c). 
On the one hand, the chosen concentration of 3.5~\% is locally above the critical concentration for pinning found for 3D distribution in the last section. 
On the other hand, it corresponds to a mean concentration of less than 0.025~\% defects in the whole system and long-range bias fields can thus be ruled out.
For all temperatures, these defects do not influence the wall before it crosses their plane, see time-evolution before the vertical dotted lines.
Particularly there is no long-range interaction.
For example at 240~K, the remaining distance between DW center and defect plane at 100~ps is only 1~u.c. 
Even for this distance, the mean wall velocity is identical to the pristine material. 
The same holds true for all tested defect concentrations, i.e.\ even planes with 7~\% defects do not show long-range coupling with the walls moving towards it, see Fig.~\ref{fig:xvr_1L3.5p}~(e)--(g).
This short-range nature of the interaction is related to the small correlation of dipoles perpendicular to the polarization direction  (cf. appendix Fig.~\ref{fig:u_corr}). 
Only nearest neighbor dipoles are weakly correlated.
Analogously, also sparse defects only couple to the polarization locally and along the polarization direction.

After the wall hits the defect plane, one can distinguish two time intervals of the wall--defect interaction for 3.5~\% defects: 
(i) pinning of wall segments which reduces the mean velocity of the wall and increases the roughness, as discussed below, and 
(ii) sequential depinning of these wall segments which restores initial shape, roughness, and velocity. 
During this process, the speed of the wall may temporarily overshoot.
At 280~K, the DWs cross the defect plane at about 10~ps and reach their maximal roughness about 20~ps later.
Afterwards the time interval (ii) \rv{exceeds} the end of the simulation
\rv{as the steady state is not fully reached at 60~ps.  Between, 30 and 60~ps the velocity of the wall exceeds the 1.7~\AA/ps found for the pristine case
and is maximal at about 40~ps \rv{as discussed below}.}
These processes are slower at 260~K as wall motion and depinning are induced by thermal fluctuations. 
At this temperature, the wall hits the defect plane at approximately 30~ps, the maximal roughness is reached at about 80~ps and interval (ii) is not yet completed after 100~ps.
Qualitatively the same trends are observed for defect concentrations up to 4~\%, see Fig.~\ref{fig:xvr_1L3.5p}~(e)--(g).
With the defect concentration, the time interval (i) and the slowing-down of the wall increase systematically. 
The maximal roughness is thus reached at 60~ps, 80~ps, or after 100~ps for 2~\%, 3.5~\%, or 4~\% defects while the minimal velocity at 50~ps is only 49~\%, 11~\%, or 4.5~\% of whose of the pristine material.
For intermediate defect concentrations, particularly for 4~\%, the DW velocity is furthermore very sensitive to the defect distribution, cf.\ dotted and dashed lines, and for more than 5~\% defects, both tested configurations fully pin the wall without sizeable changes of DW roughness. 
% mean Rdw at 100ps: 1.87~u.c. (s0), 3.53~u.c. (s1) => percentage difference = (3.53-1.87)/[(3.53+1.87)/2] = 0.61
Note that this local defect concentration corresponds to an average number of 0.03~\% defects in the chosen simulation cell.
% 5/164=0.03

For simplicity, we restrict the following detailed analysis to 3.5~\% defects and 260~K.
Figure~\ref{fig:xvr_1L3.5p}~(d) shows the underlying time-evolution of $x_0(z)$, the lines of averaged wall center positions over $y$. 
Analogous to the defect-free material, the walls are initially rather flat and move with constant macroscopic velocity but host the nucleation of 2D clusters.
At 30~ps (purple-red line) after the wall has hit the defect plane, the local probability for nucleation and growth in a large part of this plane is strongly reduced, and the wall velocity temporarily drops.
At the next shown snapshot, one cluster (red) has penetrated the defect plane at the periodic boundary (spanning from about $z=40$ to the next periodic image, $z=1$) and has grown down and up to $z=14$ and $z=9$ at 90~ps. 
During this time, no other nucleus penetrates the defect plane, and while the cluster grows along $x$ by 14~u.c., the wall segments around $z=10$ are pinned. 
In turn the wall is bending and thus its roughness increases to 3~u.c. 
Local thermal fluctuations and the local wall width are however not significantly modified. 
Finally, after 100~ps, the wall is fully depinned from the defect plane, and the wall has started to flatten again with a maximal variation of $x_0(z)$ of 11~u.c.
Within longer simulations, the steady state of the moving flat wall was finally restored. Qualitatively the same trends are found for defect planes with 2--4~\% defects.
\begin{figure}[t]
    \centering
    %\resizebox{.9\linewidth}{!}{
    \begin{Overpic}[abs]{\begin{tabular}{p{.45\textwidth}}\vspace{.188\textheight}\\\end{tabular}}
        \put(5,0){\includegraphics[height=0.2\textheight]{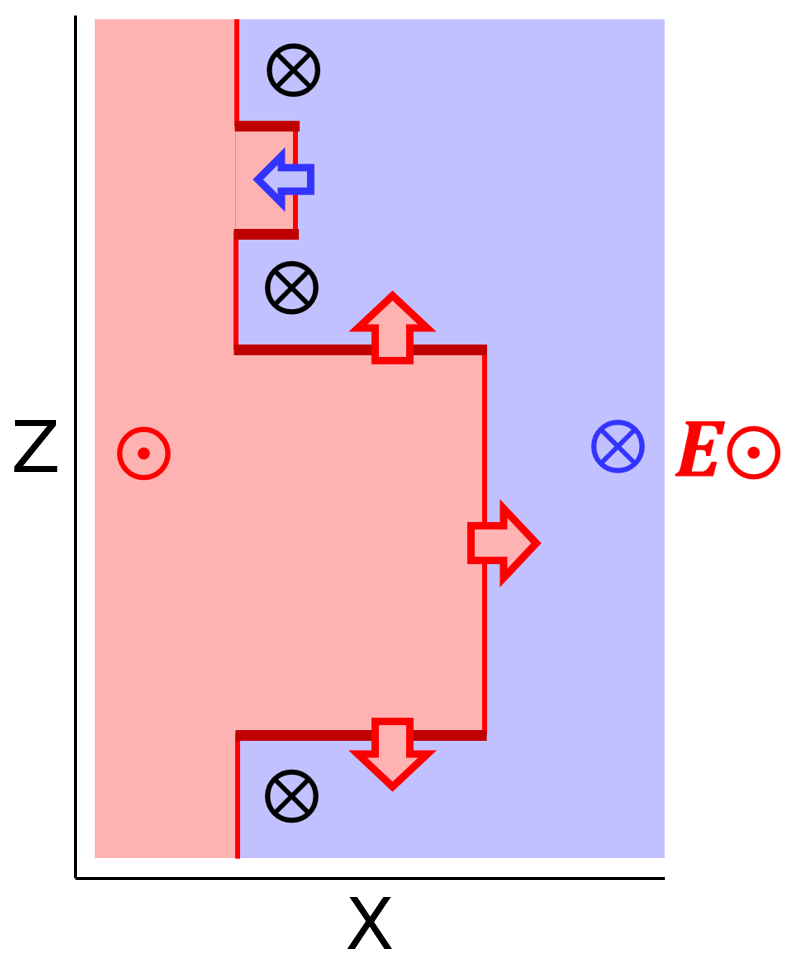}}
        \put(125,0){\includegraphics[height=0.2\textheight]{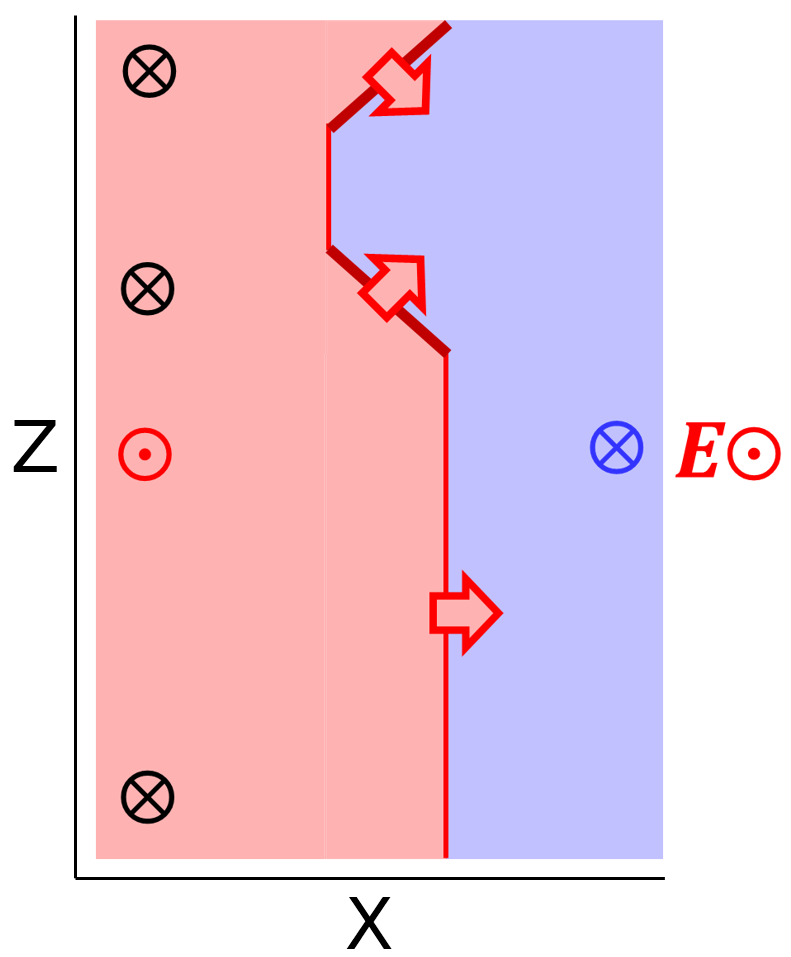}}
        \put(0,125){\textsf{\normalsize(a)}}
        \put(120,125){\textsf{\normalsize(b)}}
    \end{Overpic}
   %}
    \caption{Schematic illustration of (a) pinning, bowing, and (b) depinning of the DW with applied field $+E_y$ at defect rich $x$-plane.
    Colored and black arrowheads and areas mark the direction of the free and defect dipoles, respectively. 
    Wide arrows indicate the direction of domain growth due to nuclei that are larger and smaller than the critical size, respectively.
    Note that the sharp corners in (a) are the simplification of penetrated DWs.}
    \label{fig:depin}
\end{figure}

The underlying mechanisms for pinning and depinning are illustrated in Fig.~\ref{fig:depin}. 
As defects cannot be switched, they cannot be part of a nucleus.
Between dense defects, local dipole fluctuations are possible. 
However, appearing clusters do not reach their critical size and vanish again, see blue arrow in the upper part of Fig.~\ref{fig:depin}~(a). 
Instead, clusters nucleate and move through large enough defect-free areas as shown in the lower part of the sketch, while the rest of the wall is pinned.
This pinning of wall segments increases the wall area in the system as two additional $\pm p_y$ walls parallel to $z$ start to grow, see red arrows in Fig.~\ref{fig:depin}.
These new interfaces share the properties of the walls along $x$ by symmetry, i.e.\ they are charge-neutral, degenerate and low energy, and face the same energy barrier for motion (TiO$_2$ planes) as well as the same driving force (the field favoring $+p_y$ domains).
Note that the sharp corners of the nuclei shown in the sketch are potentially high in energy. 
Instead, the wall is locally bending and parts of it are no longer along a $\langle 100\rangle$-type direction, see Fig.~\ref{fig:xvr_1L3.5p}~(d).

The applied field induces the shift of wall segments along $x$ ($x$-walls) and $z$ ($z$-walls) with equal probability.
Therefore, the pinned segment shrinks along $z$ and the total wall velocity along $x$ increases due to the additional increase of switched dipoles behind the front of the wall. 
Therefore, the velocity can temporarily exceed that of the pristine material as shown \rv{for 260~K and 2--3.5\% defects} in Fig.~\ref{fig:xvr_1L3.5p}~(f) \rv{or for 3.5\% at 280~K in  Fig.~\ref{fig:xvr_1L3.5p}~(b).}
At the same time, the increase in energy with the increasing wall area is minor and no slowing down of the depinned wall segments becomes visible with the chosen resolution. 
For example, the velocity of the wall for the segments $z=30$--$48$~u.c.\ is identical to whose of the pristine material. 
% (1~\AA/ps)=0.25*4
By the combined wall growth along $x$ and $z$ the wall can also flow around the defect-rich region and is depinned. 
After depinning, this superimposed motion also results in the flattening of the wall as sketched in Fig.~\ref{fig:depin}~(b).

\begin{figure}[t]
    \centering
    \begin{Overpic}[abs]{\begin{tabular}{p{.16\textwidth}}\vspace{.24\textheight}\\\end{tabular}}
        \put(0,0){\includegraphics[height=0.25\textheight]{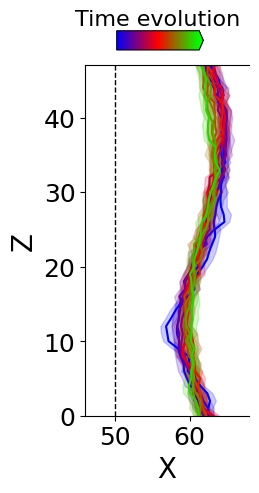}}
        \put(0,160){\textsf{\normalsize(a)}}
        \put(80,137){\color{black}\vector(-2,-1){15}}
        \put(80,90){\color{black}\vector(-2,1){15}}
        \put(57,43){\color{black}\vector(1,1){10}}
        \put(57,66){\color{black}\vector(1,-1){10}}
    \end{Overpic}
    \begin{Overpic}[abs]{\begin{tabular}{p{.15\textwidth}}\vspace{.24\textheight}\\\end{tabular}}
        \put(0,0){\includegraphics[height=0.25\textheight]{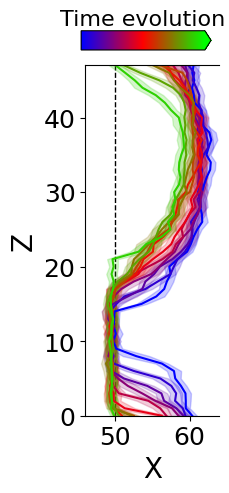}}
        \put(0,160){\textsf{\normalsize(b)}}
        \put(70,80){\color{black}\vector(-1,1){15}}
        \put(70,145){\color{black}\vector(-1,-1){15}}
        \put(47,62){\color{black}\vector(0,1){25}}
        \put(47,49){\color{black}\vector(0,-1){25}}
        \put(-18,51){\color{red}\large$\boldsymbol{E}$=0}
        \put(60,51){\color{red}\large$\boldsymbol{E}$=0}
    \end{Overpic}
    \caption{Change of the DW profile $x_0(z)$ with time for 3.5~\% \rv{defects} at 260~K \rv{after the field had been applied for (a) 100~ps (fully depinning) or (b) 90~ps (partially depinning) and was revmoved again.} 
    Dashed lines indicate the position of the defect plane and colors encode the further time evolution of 100~ps in intervals of 10~ps.}
    \label{fig:dw_restore}
\end{figure}
What happens to the DWs, if the field is removed before that?
Figure~\ref{fig:dw_restore} shows the two possible scenarios for the DW evolution after field removal during the depinning process. 
In (a) the field has been removed after 100~ps, after the wall has been detached from the defect plane but was still bended and in (b) the field has been removed from the maximally distorted wall after 90~ps.
In the former case, the DW flattens with time but stays at about $X_{\text{DW}}=62$.
Importantly, there is no attraction of the wall by the defect plane for the initial distance of about 5~u.c., and the defects can not restore the initial domain structure.
In the latter case, the wall slowly relaxes back to the defect plane. 
The defect plane thus acts as a restoring force on the initial domain structure.
Figure~\ref{fig:dw_restore}~(a) shows the time evolution of the wall if the field is removed after depinning. 
Without the external field, there is no macroscopic driving force for wall motion and the mean wall potion $X_{\text{DW}}=62$ does not change further with time. 
However, the bending decreases and the system relaxes towards a flat wall aligned along $x$. 
By that the DW area in the material is minimized and walls not along $\langle 100\rangle$-type directions, which have been predicted to be higher in energy,\cite{lawless180degDomainWallEnergies1968} vanish. 
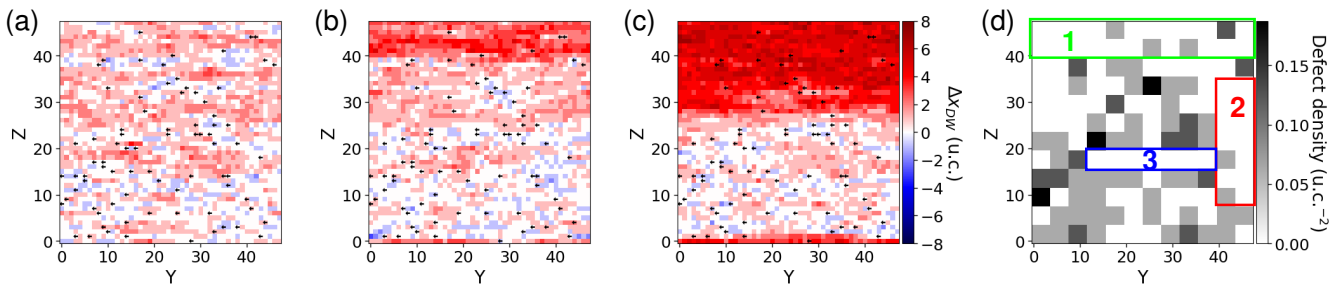
\begin{figure*}[t]
    \centering
    \resizebox{1.\linewidth}{!}{
    \begin{Overpic}[abs]{\begin{tabular}{p{.16\textwidth}}\vspace{.11\textheight}\\\end{tabular}}
        \put(0,0){\includegraphics[height=0.12\textheight,trim={0 0 17.4cm 0},clip]{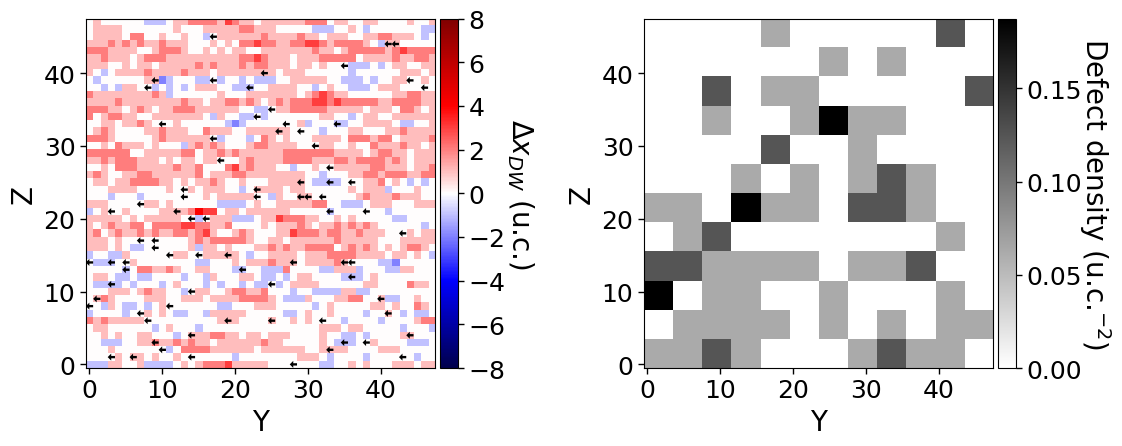}}
        \put(0,75){\textsf{\footnotesize(a)}}
    \end{Overpic}
    \begin{Overpic}[abs]{\begin{tabular}{p{.16\textwidth}}\vspace{.11\textheight}\\\end{tabular}}
        \put(0,0){\includegraphics[height=0.12\textheight,trim={0 0 17.4cm 0},clip]{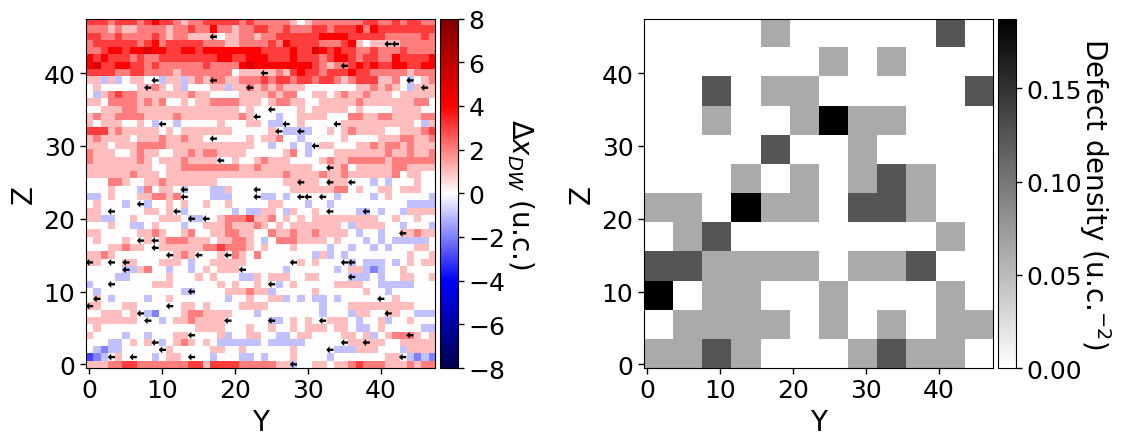}}
        \put(0,75){\textsf{\footnotesize(b)}}
    \end{Overpic}
    \begin{Overpic}[abs]{\begin{tabular}{p{.39\textwidth}}\vspace{.11\textheight}\\\end{tabular}}
        \put(0,0){\includegraphics[height=0.12\textheight]{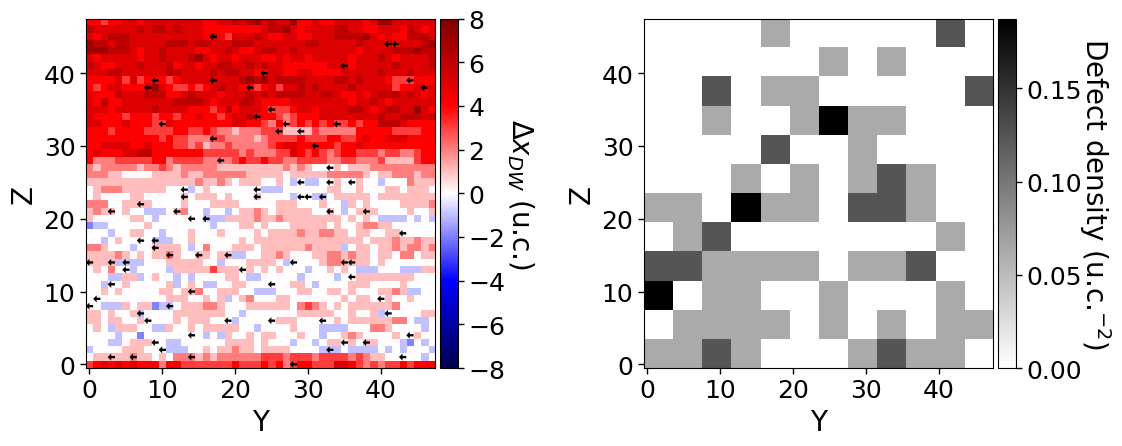}}
        \put(0,75){\textsf{\footnotesize(c)}}
        \put(102,75){\textsf{\footnotesize(d)}}
        \put(116,67){\linethickness{0.25mm}\color{green}\polygon(0,0)(64,0)(64,11)(0,11)}
        \put(125,69){\color{green}\bf\textsf{\footnotesize 1}}
        \put(169,25){\linethickness{0.25mm}\color{red}\polygon(0,0)(11,0)(11,36)(0,36)}
        \put(173,50){\color{red}\bf\textsf{\footnotesize 2}}
        \put(132,35){\linethickness{0.25mm}\color{blue}\polygon(0,0)(37,0)(37,6)(0,6)}
        \put(148,35){\color{blue}\bf\textsf{\footnotesize 3}}
    \end{Overpic}
    }
    \caption{Exemplary DW surface profile of BaTiO$_3$ at 260~K during propagation through a defect-rich layer with 3.5~\%  defects after (a) 40~ps, (b) 50~ps, and (c) 60~ps. 
    The colors describe the distance of DW segments to the defect plane and the black arrows mark the positions of defect dipoles.
    (d) The mean local defect density in $4\times 4$~u.c..
    Colored enclosed areas mark defect-free regions with large connected areas (1, 2, and 3) and regions.}
    \label{fig:dw_snap}
\end{figure*}
Note that one also may expect different energy barriers for the motion of walls along \rv{low}-symmetry directions.
If the field is however removed before full depinning, the defects still act as restoring forces on the wall, even if the mean distance between wall and defect plane along $x$ reach 8~u.c., see Fig.~\ref{fig:dw_restore}~(b).

The representation of $x_0(z)$ discussed so far in Fig.~\ref{fig:xvr_1L3.5p}~(d) is not sufficient to understand the different pinning potential of specific defect distributions. 
Although the z-slices without defects and those with a maximal local density seem to correlate with penetrated and pinned wall segments, there is no \rv{one}-to-\rv{one} correspondence as shown in appendix Fig.~\ref{fig:defects_1L3.5p_z}.
Instead the defect distribution has to be analyzed in 2D. 
Figure~\ref{fig:dw_snap}~(a)--(c) show snapshots of the local dipoles on the defect plane with 3.5~\% defects in relation to the defect distribution as shown in subfigure~(d). 
After 40~ps, the mean wall center is localized on the defect plane with local variations of $\pm$~0.86~u.c. in qualitative agreement with the typical interface of moving DWs in the pristine material. 
After 50~ps, the average wall position has already passed the defect plane by 0.4~u.c. as dipoles above $z=40$ have shifted by up to 4.5~u.c. and a 2D stripe domain has formed. 
After 60~ps, the mean wall center is already 1.7~u.c.\ in front of the defect plane, while wall segments below $z=25$ are still pinned as shown in subfigure~(c).
One can understand this segmentation of the wall into initially pinned and moving regions by the local defect distribution. 
For the chosen defect distribution, three large connected defect-free areas are marked in Fig.~\ref{fig:dw_snap}~(d). 
As we determined an approximate size of critical clusters of 4 and 12 u.c.\ along $z$ and $y$ for the pristine material, only region 1 may thus allow for the nucleation of a new 2D cluster. 
Indeed, region 3 does not switch the sign of polarization within the first 60~ps, neither does region 2 host a new nucleus.
In order to estimate the critical size of a defect-free region for the wall penetration, we analyze defect planes with equidistant slits along $y$ and $z$.
The DW can cross these defect-free stripes if they are larger than a critical width of 4 and 40~u.c., respectively (cf.\ appendix Fig.~\ref{fig:dw_slit}).
Perpendicular to the polarization direction,  it is thus sufficient to have a defect-free slit with the width of the critical nucleus of the pristine material.
The defect concentration can exceed 20~\% and the wall is still not pinned.
This underlines the short-range interaction of DWs and defects perpendicular to the polarization direction, as already discussed for the $x$-direction above.
Only along the direction of the polarization $y$, the ``local bias'' fields induced by the defects increase the critical size of a defect-free slit compared to the critical nucleus by a factor of about 4.
This short-range and anisotropic coupling between defects and wall also explains the large spread of wall velocities between independent runs for intermediate defect concentrations (135~\% for 4~\% defects).
% 2/3.5\%: 0.18, 0.19~\AA/ps (21--34~\%); 4\%: 0.28, 0.32~\AA/ps (51--135~\%)

In summary, neither macroscopic nor local defect concentrations determine the pinning of DWs, but rather the dilation of connected defect-free regions on the wall interface is decisive. 
Agglomerated defects with sufficiently large holes thus cannot pin the wall, even if their concentration reaches about 20~\%. 
Thereby, the defects mainly influence the dipoles along the $y$-direction.

%%%%%%%%%%%%%%%%%%%%%%%%%%%%%%

\section{Conclusion}
\label{sec:conclusion}

We analyzed the impact of acceptor doping and defect dipoles on the field-induced motion of 180$^{\circ}$ DWs in the tetragonal phase of BaTiO$_3$ by \textit{ab\ initio} derived molecular dynamics simulations.
We could verify the predictions by Ref.~\onlinecite{renLargeElectricfieldinducedStrain2004} that such defects if aligned with the local polarization after aging, act as local internal restoring forces on the domain structure. 
For moderate field strengths, we find a critical concentration of 2~\% defects randomly distributed in the sample to pin the DWs. 

\rv{Note that this absolute value of the critical concentration and all quantitative predictions in our work have a sizeable errorbar due to the used approximation of frozen-in defect dipoles and the neglected strain relaxation around the defects, which are beyond the scope of the current work. In order to address defect dynamics, e.g., in large field strengths, or the strain coupling, e.g., in case of elastic domain walls, future work can combine the effective Hamiltonian with kinetic Monte Carlo simulations or add local strain fields to improve the description, respectively.}

\rv{Importantly}, our simulations give new insights in the coupling mechanisms between defects and inelastic DWs:
The interaction between defect dipoles and these walls is surprisingly short-ranged and anisotropic.
Therefore, not only defect density and type of defects, but particularly the defect distribution governs their pinning potential.
Walls can flow around defects and defect clusters by thermally activated motion if the defect distribution offers defect-free areas in front of the moving wall which are larger than the critical nucleus. 
Perpendicular to the polarization direction, the critical spacing is about 1.6~nm only and even 20~\% of defects could not pin the wall if defect-free slits \rv{exist}.

The fact that the wall--defect interaction is thus very sensitive to the defect distribution, has two important consequences for the functional properties of the material: 
First, in inhomogeneous samples or if the defect distribution changes with time due to diffusion and segregation, the wall dynamics and thus the local polarization switching may vary drastically. 
Second, it is possible to control the wall motion on the nanoscale by properly designing the defect distribution. 
Particularly, confining defects in 2D areas provides a maximal pinning potential. 
Furthermore, the anisotropy of the defect--wall interaction is promising to control wall bending and regulate the direction of wall motion.

%%%%%%%%%%%%%%%%%%%%%%%%%%%%%%

\begin{acknowledgments}
This work was supported by the German research foundation (DFG) GR 4792/3. 
\end{acknowledgments}

\section*{Data Availability Statement}
The data that support the findings of this study are openly available in Zenodo repository at \url{https://zenodo.org/records/14673466}, Ref.~\onlinecite{tengControlFerroelectricDomain2025}.
All analysis scripts associated with this work are freely available in a Gitlab repository at \url{https://gitlab.ruhr-uni-bochum.de/tengssh/fedas}, Ref.~\onlinecite{tengShengHanTengFedas2025}.

%%%%%%%%%%%%%%%%%%%%%%%%%%%%%%

\appendix

\section{Defect dipoles and wall motion: Random distribution}
\begin{figure}[h]
    \centering
    %\resizebox{.9\linewidth}{!}{
    \begin{Overpic}[abs]{\begin{tabular}{p{.4\textwidth}}\vspace{.29\textheight}\\\end{tabular}}
        \put(0,0){\includegraphics[width=0.4\textwidth]{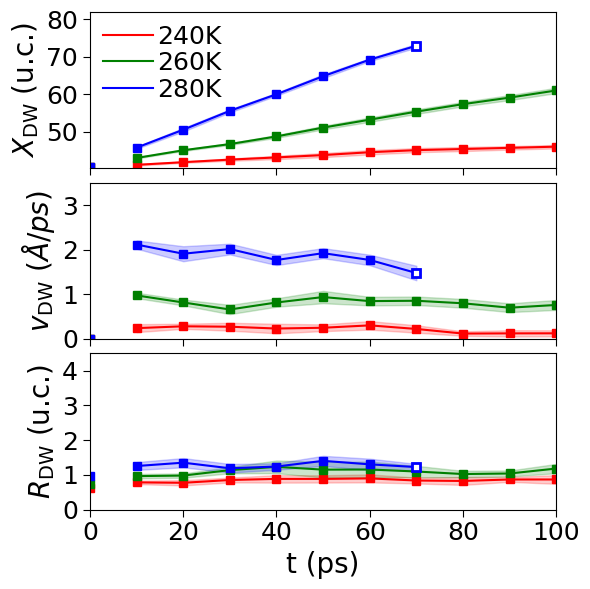}}
        \put(178,188){\textsf{\normalsize(a)}}
        \put(178,128){\textsf{\normalsize(b)}}
        \put(178,68){\textsf{\normalsize(c)}}
    \end{Overpic}
    %}
    \caption{Temperature dependence of DW (a) position,  $X_{\text{DW}}$, (b) velocity,  $v_{\text{DW}}$, and (c) roughness, $R_{\text{DW}}$, in the preseence of 0.5~\% randomly distributed frozen-in defect dipoles of $u_d=\pm 0.09$~\AA~$\hat{y}$ aligend with the inital polarization direction at 240~K (red), 260~K (green) and 280~K (blue).
    At each temperature, the mean values (solid line) and standard deviation (shaded area) of 10 independent simulations are shown.
    At 280~K, both DWs get too close for further analysis after 70~ps (empty squares) in the chosen simulation cell size of 164~u.c.}
    \label{fig:xvr_r3d.5p}
\end{figure}
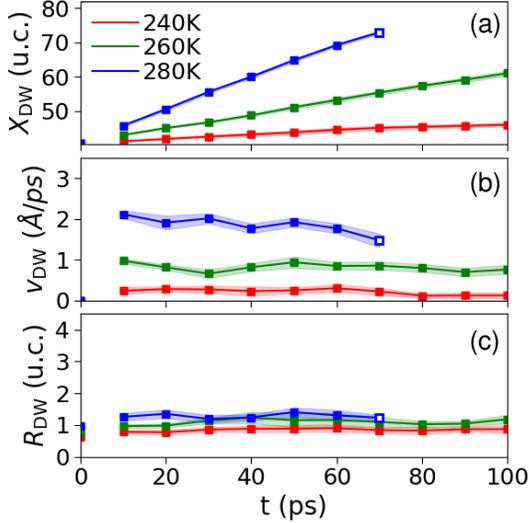

This section collects additional data on DW motion in the presence of randomly distributed defects: 
As shown in Figure~\ref{fig:xvr_r3d.5p} for 0.5~\%  defects, the temperature dependency of the wall motion in the presence of a small defect concentration is similar to the pristine material, cf.~Fig.~\ref{fig:xvr_pristine}. 
The main difference is the suppression of the initial wall acceleration by non-equilibrium effects.

\begin{figure}[t]
    \centering
   \resizebox{\linewidth}{!}{
	  \begin{Overpic}[abs]{\begin{tabular}{p{.42\textwidth}}\vspace{.142\textheight}\\\end{tabular}}
        \put(0,0){\includegraphics[height=0.15\textheight,clip=true,trim=0cm 0cm 3cm 0cm]{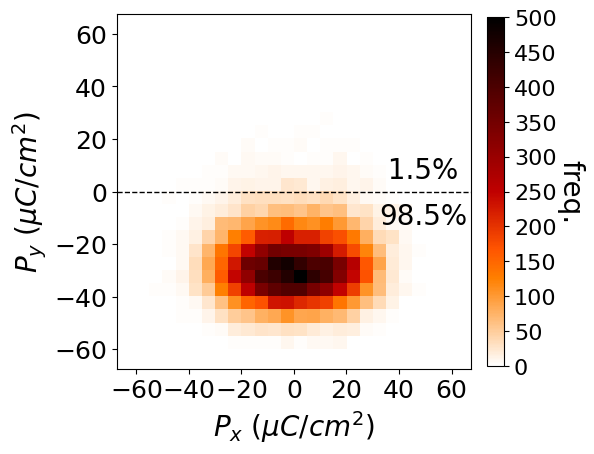}}
        \put(29,88){\textsf{\normalsize(a)}}
        \put(110,0){\includegraphics[height=0.15\textheight,clip=true,trim=2.7cm 0cm 0cm 0cm]{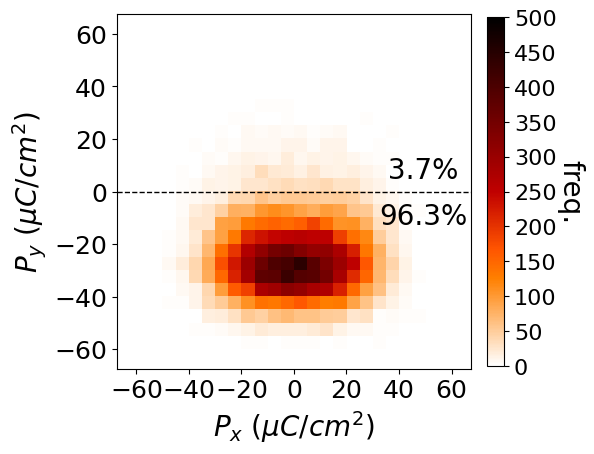}}
        \put(114,88){\textsf{\normalsize(b)}}
    \end{Overpic}
	  }
    \caption{\rv{Initial non-equilibrium switching of dipoles in the middle of $-p_y$ domain ($x=82$) after the field ($E_y=+100$~kV/cm) has been applied for 10~ps for pristine BaTiO$_3$ at (a) 260~K and (b) 280~K.
    Colors encode the cumulated number of dipoles per (5 $\mu$C/cm$^2$)$^2$ area of 10 independent simulations.
    The dipole distribution on z-plane is approximately the same as on x-plane (not shown).}}
    \label{fig:uhist_260_280K}
\end{figure}

\rv{Figure~\ref{fig:uhist_260_280K} visualizes how the initial non-equilibrium switching of dipoles in the $-p_y$ domain after fast field application changes with temperature.
The fraction of non-equilibrium dipoles after field application for 10~ps is higher at 280~K than at 260~K.
As discussed in Ref.~\onlinecite{khachaturyanDomainWallAcceleration2022} these initially switched dipoles promote the nucleation and growth underlying domain wall motion and thus boost the domain wall velocity. 
With time (not shown), the dipoles switch back and the domain wall velocity reaches its steady state.}\\

\begin{figure}[t]
    \centering
    %\resizebox{.9\linewidth}{!}{
    \begin{Overpic}[abs]{\begin{tabular}{p{.4\textwidth}}\vspace{.29\textheight}\\\end{tabular}}
        \put(0,0){\includegraphics[width=0.4\textwidth]{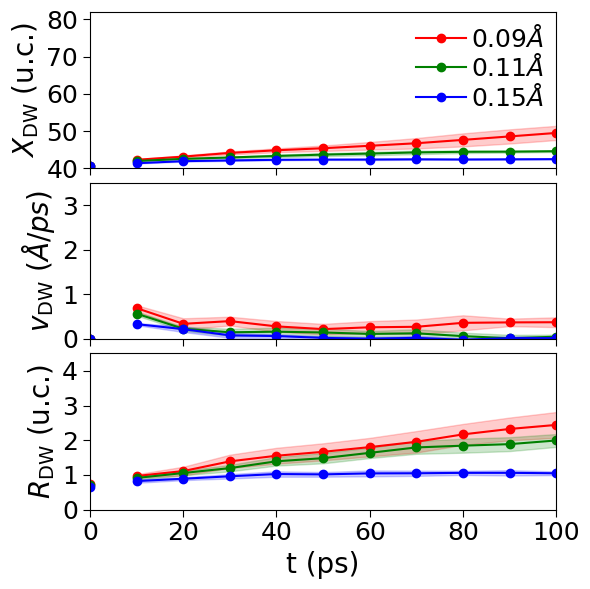}}
        \put(35,188){\textsf{\normalsize(a)}}
        \put(35,128){\textsf{\normalsize(b)}}
        \put(35,68){\textsf{\normalsize(c)}}
    \end{Overpic}
    %}
    \caption{Impact of defect dipole strength on the time evolution of DW (a) position, $X_{\text{DW}}$, (b) velocity, $v_{\text{DW}}$, and (c) roughness, $R_{\text{DW}}$, for  1~\% randomly distributed frozen-in defect dipoles corresponding to $u_d=\pm 0.09$ (red), $\pm 0.11$ (green), and $\pm 0.15$~\AA (blue) at  260~K.
    Mean values (solid line) and standard deviation (shaded area) of 10 independent simulations are given.}
    \label{fig:xvr_r3d1p_ud}
\end{figure}

Figure~\ref{fig:xvr_r3d1p_ud} shows the impact of the defect dipole strength on the wall motion for the example of 1~\% defects at 260~K comparing $u_d=0.09$ ($0.8\cdot P_y$), $0.11$ ($0.98\cdot P_y$), and $0.15$~\AA ($1.33\cdot P_y$).
With the defect strength, the wall velocity decreases and after 80~ps the wall position is already pinned for 1~\% defects for stronger defects.

\begin{figure}[h]
    \centering
    \begin{Overpic}[abs]{\begin{tabular}{p{.4\textwidth}}\vspace{.41\textheight}\\\end{tabular}}
        \put(0,140){\includegraphics[width=0.4\textwidth]{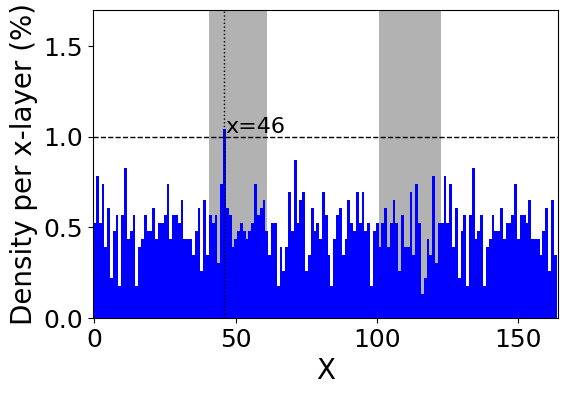}}
        \put(0,0){\includegraphics[width=0.4\textwidth]{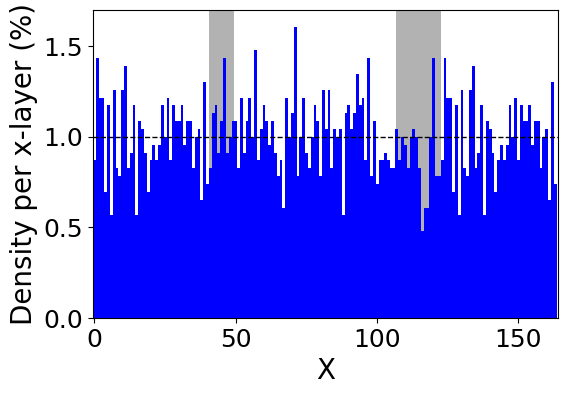}}
        \put(35,268){\textsf{\normalsize(a)}}
        \put(35,128){\textsf{\normalsize(b)}}
    \end{Overpic}
    \caption{Local defect density per $x$-layer for \rv{(a) 0.5~\% and (b) 1}~\% randomly distributed defects. \rv{In (a) the local defect density exceeds 1~\% at $x=46$.
    The gray areas mark the shift of the domain walls at 260~K within 100~ps in a field of $E_y=+100$~kV/cm for walls moving from $x=40.5$ to the right and from $x=122.5$} to the left.}
    \label{fig:defects_r3d.5p_x}
\end{figure}

Figure~\ref{fig:defects_r3d.5p_x} shows the layer-resolved defect density \rv{for the defect configurations used in  Sec.~\ref{ssec:r3d} for (a) 0.5~\% and (b) 1~\% randomly distributed defects}.
\rv{The local defect concentrations in each plane deviate from their global mean. Particularly, the local concentration in layer $x=46$ in case (a) even exceeds 1~\%. 
The gray areas in the figure indicate the layers crossed by the moving walls during 100~ps. Note that each simulation cell contains two domain walls, the one moving to the right discussed in the main text and one moving to the left initially at $x=122.5$. 
Due to the different defect distributions in front of these walls, the number of crossed layers differ between left and right walls. 
For 0.5~\%, the interaction between defects and wall is small, however, due to the one layer with larger concentration, the traveled distance is 1~u.c. less for the left wall. 
For 1\% defect, the right wall moves 7~u.c. more as it crosses a region with less defects.}

\newpage

\section{Defect plane}

\begin{figure}[t]
    \centering
    %\resizebox{.9\linewidth}{!}{
    \begin{Overpic}[abs]{\begin{tabular}{p{.38\textwidth}}\vspace{.188\textheight}\\\end{tabular}}
        \put(0,0){\includegraphics[height=0.2\textheight]{wl_1L3.5p_s0_0.png}}
        \put(90,0){\includegraphics[height=0.181\textheight]{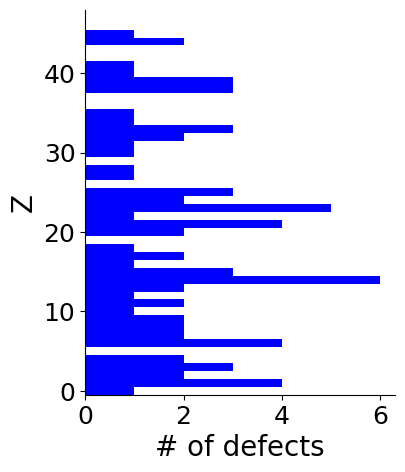}}
        \put(0,123){\textsf{\normalsize(a)}}
        \put(90,123){\textsf{\normalsize(b)}}
	\end{Overpic}
	%}
   \caption{(a) Change of the DW profile $x_0(z)$ with time for 3.5~\% defects on the plane $x=50$ at 260~K, cf. Fig.~\ref{fig:xvr_1L3.5p}.
   Colors encode the time after field application from 0~ps (blue) to 100~ps (green) in intervals of 10~ps.
   (b) Underlying $z$-profile of the number of defects.}
   \label{fig:defects_1L3.5p_z}
\end{figure}

Figure~\ref{fig:defects_1L3.5p_z} compares the difference of the evolution of the wall positions along $z$ ($x_0(z)$) from Fig.~\ref{fig:xvr_1L3.5p} with the corresponding aggregated defect densities. 
While the maximal and minimal numbers of defects are close to the pinned segment around $z=10$ and the segment propagating through the defect plane first, there is no one-to-one correspondence. 

\section{Toy slit model}

\begin{figure}[h]
    \centering
    %\resizebox{.9\linewidth}{!}{
    \begin{Overpic}[abs]{\begin{tabular}{p{.2\textwidth}}\vspace{.11\textheight}\\\end{tabular}}
        \put(0,0){\includegraphics[height=0.12\textheight]{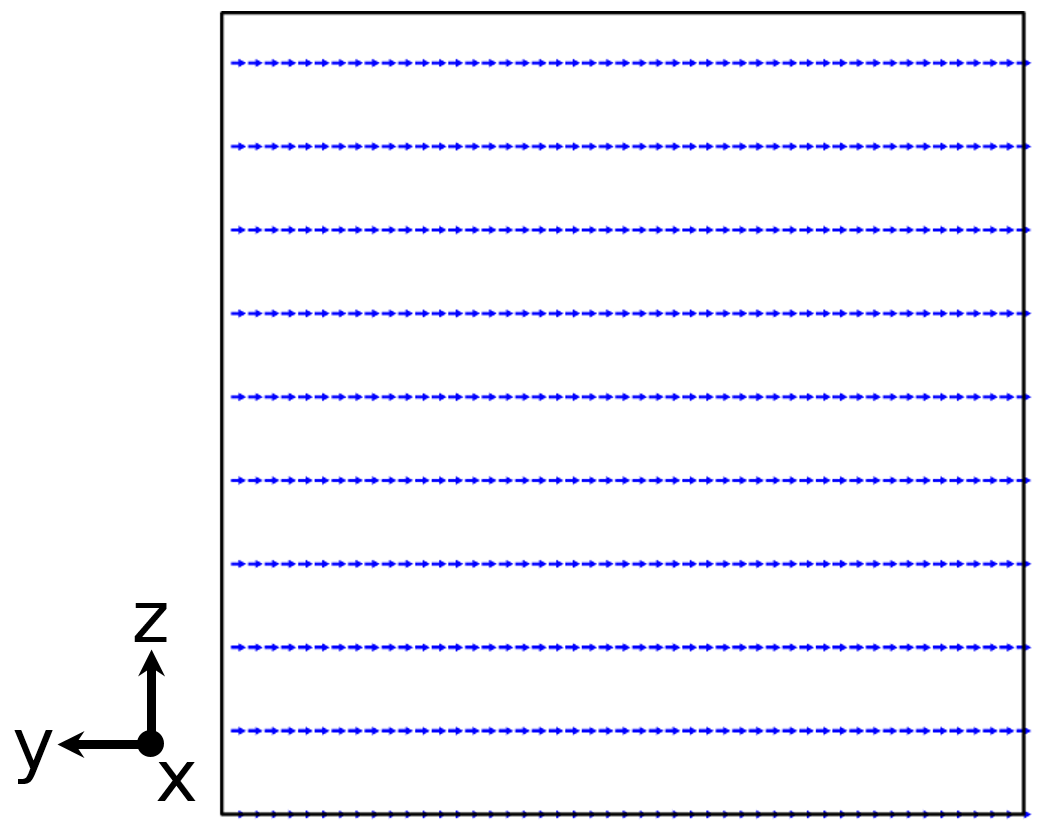}}
        \put(0,75){\textsf{\normalsize(a)}}
    \end{Overpic}
    \begin{Overpic}[abs]{\begin{tabular}{p{.2\textwidth}}\vspace{.11\textheight}\\\end{tabular}}
        \put(0,0){\includegraphics[height=0.12\textheight]{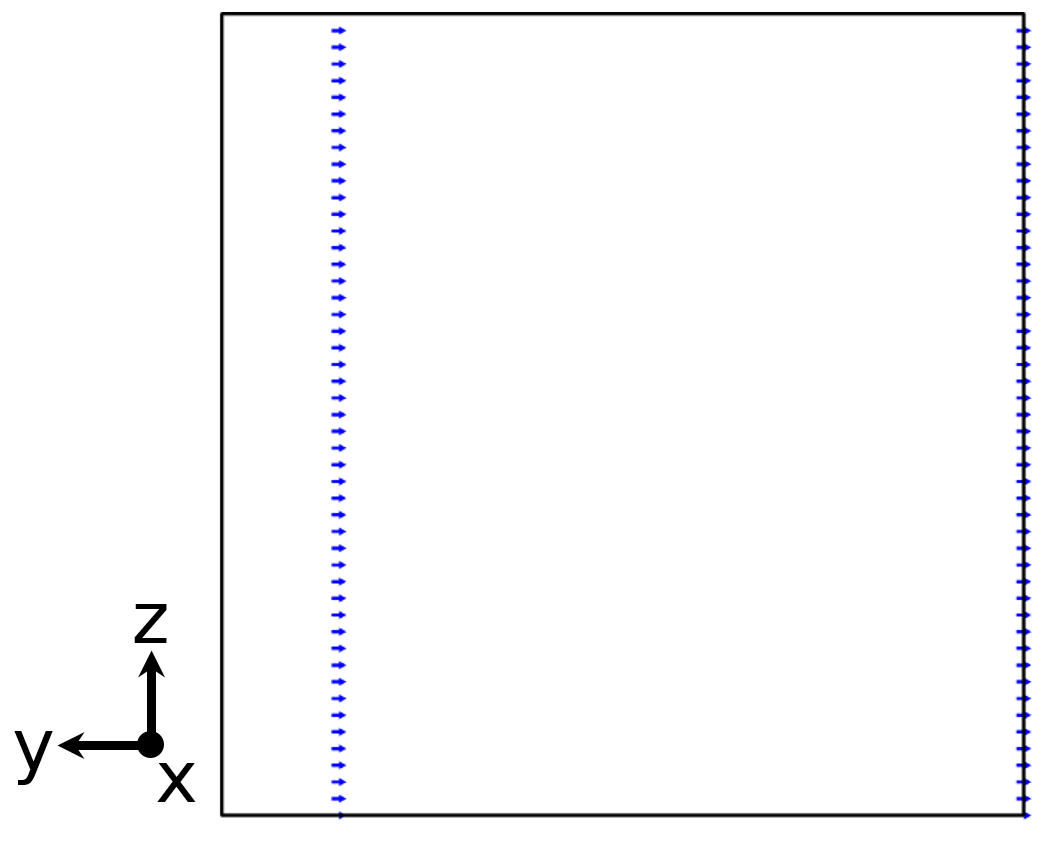}}
        \put(0,75){\textsf{\normalsize(b)}}
    \end{Overpic}
     \begin{Overpic}[abs]{\begin{tabular}{p{.2\textwidth}}\vspace{.24\textheight}\\\end{tabular}}
        \put(0,0){\includegraphics[height=0.25\textheight]{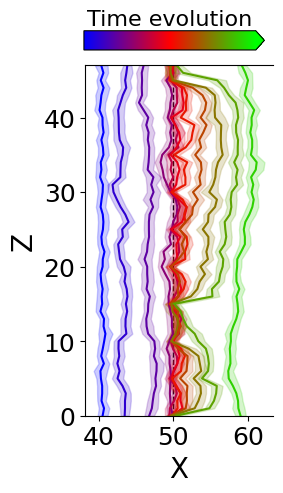}}
        \put(0,157){\textsf{\normalsize(c)}}
    \end{Overpic}
    \begin{Overpic}[abs]{\begin{tabular}{p{.21\textwidth}}\vspace{.24\textheight}\\\end{tabular}}
        \put(0,0){\includegraphics[height=0.25\textheight]{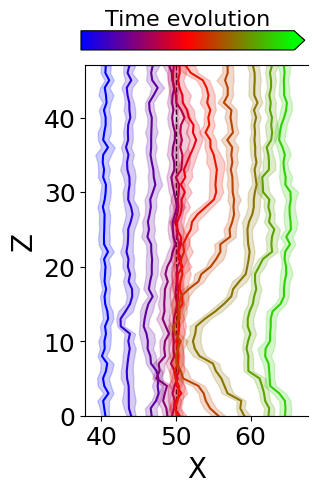}}
        \put(0,157){\textsf{\normalsize(d)}}
    \end{Overpic}
    %}
    \caption{DW propagation through a plane with defect lines and slits with (a) defect-free areas of $z=4$~u.c. and (b) defect-free areas of $y=40$~u.c. (c) Penetration of the wall through the defect plane shown in (a).  (d) Penetration of the wall through the defect plane shown in (b).}
    \label{fig:dw_slit}
\end{figure}

To test the hypothesis of critical nuclei for domain nucleation and growth, we set up 2 toy systems with defects in slit arrangements as shown in Fig.~\ref{fig:dw_slit}~(a)--(b).
Figure~\ref{fig:dw_slit}~(c) shows that despite a local defect density of 20~\%, the wall can cross a plane with slits with a width of $z=4$~u.c.
At all slits, the nucleation and growth set in with equal probability, with time the depinned areas merge, and finally the wall is depinned. 
As shown in (d), a plane with slits of widths $y=40$~u.c. can also be crossed by the wall. 

\newpage

\section{Dipole-dipole correlation}

\begin{figure}[h]
    \centering
    \begin{Overpic}[abs]{\begin{tabular}{p{.4\textwidth}}\vspace{.23\textheight}\\\end{tabular}}
        \put(0,0){\includegraphics[height=0.24\textheight]{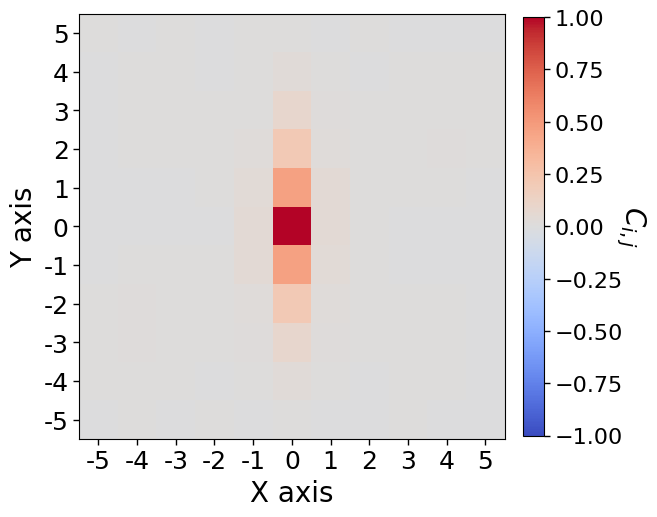}}
        \put(0,150){\textsf{\normalsize(a)}}
	\end{Overpic}
	\begin{Overpic}[abs]{\begin{tabular}{p{.4\textwidth}}\vspace{.23\textheight}\\\end{tabular}}
        \put(0,20){\includegraphics[width=0.38\textwidth]{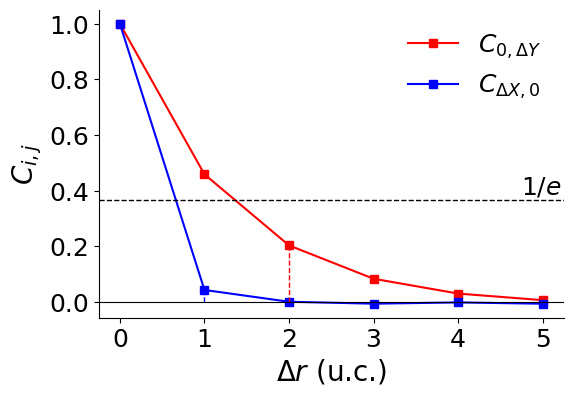}}
        \put(0,150){\textsf{\normalsize(b)}}
	\end{Overpic}
    \caption{(a) Exemplary two-point correlation map of $p_y$ in the $+p_y$ domain of the pristine material. 
    Colors encode the magnitude of the correlation coefficient $C_{i,j}$.
    (b) The cross-cut along $+\Delta X$ (blue) and $+\Delta Y$ (red). 
    Vertical dashed lines indicate the critical length below the magnitude of $1/e$.}
    \label{fig:u_corr}
\end{figure}

This section collects additional data on the dipole-dipole correlations.
The two-point polarization correlation is given as
\begin{equation}
    C_{i,j}=\frac{\langle p_y(\bm{r}_i)p_y(\bm{r}_j)\rangle-\langle p_y(\bm{r}_i)\rangle \langle p_y(\bm{r}_j)\rangle}{\sigma_i \cdot \sigma_j},
\end{equation}
where $\bm{r}_i$ and $\bm{r}_j$ are the positions of the dipoles  and $\sigma_i$ and $\sigma_j$ are the corresponding standard deviations of the given dataset.
Figure~\ref{fig:u_corr}~(a)--(b) show the two-point polarization correlations in the $+p_y$ domain of the pristine material.
Only along the polarization direction $y$, the correlation is larger than $0.2$ for distances up to 2~u.c. 
Perpendicular to it, it is short-ranged and only $0.043$ for nearest neighbors. 
Note that only $x$-direction is shown as the $z$-direction is identical per symmetry.

%%%%%%%%%%%%%%%%%%%%%%%%%%%%%%

\clearpage
\bibliographystyle{apsrev} % apsrev, apsrmp, abbrv, alpha, acm, apalike, ieeetr, plain, unsrt
\bibliography{references.bib}

\end{document}